\documentclass[fleqn,usenatbib]{mnras}

\usepackage{newtxtext,newtxmath,ulem}

\usepackage[T1]{fontenc}

\DeclareRobustCommand{\VAN}[3]{#2}
\let\VANthebibliography\thebibliography
\def\thebibliography{\DeclareRobustCommand{\VAN}[3]{##3}\VANthebibliography}

\usepackage{graphicx}	
\usepackage{amsmath}	
\usepackage{subcaption}
\usepackage{tabularx} 
\usepackage{booktabs}  
\usepackage{xcolor}
\usepackage{arydshln}
\usepackage{multirow}
\usepackage{array}
\usepackage{orcidlink}
\usepackage{xspace}
\usepackage{booktabs}
\usepackage[dvipsnames]{xcolor}

\newcommand{\change}[1]{#1}
\newcommand{\HI}{H\,\textsc{i}\xspace}

\title[\HI linewidth measurement]{WALLABY pilot survey: accurate and automated linewidth measurements from global \HI profiles}

\author[Shinna Kim et al.]{Shinna Kim$^{1,2}$\orcidlink{0000-0003-3143-5624}\thanks{E-mail: shinna.kim@research.uwa.edu.au},
Tobias Westmeier$^{1,2}$\orcidlink{0000-0002-5300-2486},
Lister Staveley-Smith$^{1, 2}$\orcidlink{0000-0002-8057-0294},
Danail Obreschkow$^{1,2}$\orcidlink{0000-0002-1527-0762},
\newauthor
Barbara Catinella$^{1,2}$\orcidlink{0000-0002-7625-562X}, 
Nathan Deg$^{3}$\orcidlink{0000-0003-3523-7633},
Karen Lee-Waddell$^{4,5}$\orcidlink{orcid.org/0000-0003-4844-8659},
Jonghwan Rhee$^{5,1}$\orcidlink{0000-0001-8496-4306},
Kristine Spekkens$^{3}$\orcidlink{0000-0002-0956-7949},
\newauthor
O. Ivy Wong$^{5,1}$\orcidlink{0000-0003-4264-3509},
Matthew Colless$^{6}$\orcidlink{0000-0001-9552-8075},
Hélène M. Courtois$^{7}$\orcidlink{0000-0003-0509-1776},
Boyi Ding$^{6}$\orcidlink{0009-0007-6458-1712},
Bumhyun Lee$^{8}$\orcidlink{0000-0002-3810-1806}
\\
$^{1}$International Centre for Radio Astronomy Research (ICRAR), The University of Western Australia, 35 Stirling Hwy, Crawley, WA 6009, Australia\\
$^{2}$ARC Centre of Excellence for All Sky Astrophysics in 3 Dimensions (ASTRO 3D)\\
$^{3}$Department of Physics, Engineering Physics and Astronomy, Queen's University, 64 Bader Lane, Kingston, Ontario, K7L 3N6, Canada \\
$^{4}$Australian SKA Regional Centre (AusSRC) - The University of Western Australia, 35 Stirling Highway, Crawley WA 6009, Australia \\
$^{5}$Australia Telescope National Facility, CSIRO Space \& Astronomy, P.O. Box 1130, Bentley, WA 6102, Australia \\
$^{6}$Research School of Astronomy and Astrophysics, Australian National University, Canberra, ACT, Australia \\
$^{7}$Universit\'e Claude Bernard Lyon 1, IUF, IP2I Lyon, 4 rue Enrico Fermi, 69622 Villeurbanne, France \\
$^{8}$Department of Astronomy, Yonsei University, 50 Yonsei-ro, Seodaemun-gu, Seoul 03722, Republic of Korea
}
\date{Accepted XXX. Received YYY; in original form ZZZ}

\pubyear{\the\year{}}

\begin{document}
\label{firstpage}
\pagerange{\pageref{firstpage}--\pageref{lastpage}}
\maketitle

\begin{abstract}
Current-generation 21-cm surveys of extragalactic neutral hydrogen ({\HI}) with instruments such as FAST and ASKAP are providing detailed spectral and spatial information for over $10^5$ galaxies. Future \HI surveys with next-generation radio telescopes could provide sample sizes larger by orders of magnitude. There is therefore a need for accurate, automated, and reliable parameterisation of the \HI emission from galaxies. We have developed an automated estimator for \HI global profile widths to better enable studies of, for example, dynamical masses and Tully-Fisher distances. Our estimator is based on an implementation of the BusyFit profile-fitting algorithm, and converges for 95\% of global profiles in the WALLABY pilot survey. Probability density functions are calculated for all profile parameters using Monte Carlo techniques, and example error parameterisations are given. We confirm that linewidth uncertainties depend primarily on the integrated signal-to-noise ratio, profile steepness, and the linewidth, consistent with previous studies. We compare several common linewidth definitions, and the width at 50\% of the mean flux density ($\ensuremath{W_{m50}}$) is found to be the most robust measure of linewidth. We provide examples of the baryonic Tully-Fisher relation drawn from the WALLABY pilot survey.

\end{abstract}

\begin{keywords}
radio lines: galaxies --
galaxies: disc --
line: profiles --
galaxies: fundamental parameters -- 
methods: data analysis
\end{keywords}

\section{Introduction}
\label{sec:introduction}

Rotational velocity is a fundamental property of disk galaxies, intimately linked to their dynamical mass via the virial theorem  \citep{2008gady.book.....B, 2002A&A...385..816D, 2008AJ....136.2761O} and to their total stellar and baryonic masses via scaling relations such as the Tully-Fisher and Fall relations \citep{1977A&A....54..661T, 1983IAUS..100..391F}. Measurements of rotation velocities are therefore essential for studying these scaling relations. Changes in these relations over cosmic time and departures from these relations due, for example, to environment, interactions, and outflows provide key constraints on galaxy evolution. 

The rotational velocity of a galaxy is primarily measured in two ways: through the linewidths of \HI global profiles; or via spatially resolved rotation curves. Historically, most measurements have relied on single-dish radio observations of (largely) unresolved galaxies, where the linewidth of the global \HI profile, provided that an accurate inclination correction can be applied, can be a good proxy for the rotational velocity in the outer `flat' part of the rotation curve. Large sky area single-dish surveys, such as the \HI Parkes All Sky Survey \citep[HIPASS;][]{2001MNRAS.322..486B} and the Arecibo Legacy Fast ALFA \citep[ALFALFA;][]{2005AJ....130.2598G, 2018ApJ...861...49H} have produced extensive datasets using this method. Over the past several decades, advances in radio interferometry instrumentation, sensitivity, and computing capabilities have enabled spatially resolved \HI mapping of increasing numbers of galaxies. Correspondingly, a range of automated kinematic modelling tools have been developed to derive rotation curves from resolved \HI data cubes and velocity fields using tilted-ring modelling techniques, including the Fully Automated TiRiFiC \citep[FAT;][]{2015MNRAS.452.3139K}, 3DBarolo \citep{2015MNRAS.451.3021D}, and 2DBAT \citep{2018MNRAS.473.3256O}. Interferometric surveys have been extensively used to map the \HI distribution and kinematics of galaxies, allowing measurements of rotation curves out to large radii and providing constraints on the mass distribution of both baryonic and dark matter (e.g., THINGS\footnote{The \HI Nearby Galaxy Survey}; \citealt{2008AJ....136.2648D}, LITTLE THINGS\footnote{Local Irregulars That Trace Luminosity Extremes, The \HI Nearby Galaxy Survey}; \citealt{2015AJ....149..180O}, LVHIS\footnote{The Local Volume \HI Survey}; \citealt{2018MNRAS.478.1611K, 2025Ap&SS.370..129O}, MHONGOOSE\footnote{MeerKAT \HI Observations of Nearby Galactic Objects: Observing Southern Emitters}; \citealt{2024A&A...688A.109D}). In parallel, optical spectroscopy using long slits, multiple fibers and integral field units (IFUs) also contribute spatially resolved measurements of a large number of galaxies. Notable recent IFU studies of large samples include SAMI\footnote{Sydney-AAO Multi-object Integral-field spectrograph}, MaNGA\footnote{Mapping Nearby Galaxies at Apache Point Observatory}, and Hector \citep{2012MNRAS.421..872C, 2015ApJ...798....7B, 2016SPIE.9908E..1FB}. However, these optical observations are often limited to the bright parts of the stellar disk and have lower spectral resolution compared to \HI (e.g., \citealt{1996ApJS..107...97M, 2023MNRAS.519.1098C}).

With the advent of next-generation radio facilities, the field is now transitioning into an era of unprecedented large-area, wide-field \HI surveys. These modern surveys are designed to scale up extragalactic samples by orders of magnitude; for example, the Widefield ASKAP L-band Legacy All-sky Blind surveY \citep[WALLABY;][]{2020Ap&SS.365..118K} is expected to detect \HI\ in approximately 200,\,000 galaxies, while the FAST all-sky \HI\ survey \citep[FASHI;][]{2024SCPMA..6719511Z} will further expand the available sample. Looking further ahead, the upcoming Square Kilometre Array \citep[SKA;][]{2009IEEEP..97.1482D} may also increase the number of available \HI profiles and resolved rotation curves in wide area surveys. In this regime, fast, robust, and automated measurement techniques are required to process large samples efficiently while maintaining consistency and minimising systematic biases.

Although surveys such as WALLABY can successfully resolve and derive rotation curves for a subset of their samples (e.g., \citealt{2022PASA...39...59D, 2024PASA...41...88M, 2024ApJ...976..159D}), the vast majority of detections in these massive surveys will inevitably remain poorly resolved. Consequently, global linewidth measurements remain highly advantageous due to their independence from spatial resolution and high computational efficiency, making them central to large statistical studies of the Tully-Fisher relation \citep{1997AJ....113...53G, 2005ApJS..160..149S, 2006ApJ...653..861M, 2011MNRAS.414.2005C, 2012MNRAS.420.1959C, 2023ApJ...950...87B}. In this context, it is essential to understand how different automated linewidth definitions introduce systematic offsets with respect to intrinsic kinematic quantities (such as the flat part of the rotation curve or the maximum rotational velocity). Previous studies have shown that linewidth measurements depend not only on the adopted measurement algorithm but also on the shape and radial extent of the rotation curve, leading to systematic differences between linewidth estimators. These effects directly propagate into the scatter and slopes of inferred scaling relations \citep{2001ApJ...563..694V, 2007AJ....134..334C, 2009AJ....138.1938C, 2016MNRAS.463.4052P, 2016ApJ...832...11B, 2019MNRAS.484.3267L}. Resolved rotation curves from interferometry therefore serve as the ultimate benchmark to validate and calibrate these global linewidth estimators.

While several automated tools have been developed to address these labor-intensive measurement approaches -- including BusyFit \citep{2014MNRAS.438.1176W}, SoFiA (\citealt{2015MNRAS.448.1922S, 2021MNRAS.506.3962W}), PANDISC \citep{2023ApJ...950..163P}, and the ALFALFA curve-of-growth pipeline \citep{2023ApJ...950...87B} -- automation alone does not necessarily reduce biases and errors. It is known, for example, that SoFiA struggles with the accurate parameterisation of linewidths at the 50\% level ($\ensuremath{W_{50}}$) for low signal-to-noise ratio spectra \citep{2021MNRAS.506.3962W}.

To measure and address the limitations of current algorithms, this paper firstly introduces the WALLABY pilot survey, which provides a set of example global spectra covering a range of signal-to-noise ratios (Section~\ref{sec:data}). We then describe our methodology, including robust error estimation (Section~\ref{sec:methods}). A comparison between the various width estimation algorithms is then made (Section~\ref{sec:results}). An application to the baryonic Tully-Fisher relation is discussed in Section~\ref{sec:discussion}, and an overall summary and conclusion are given in Section~\ref{sec:summary}. Throughout this paper, we adopt a flat $\Lambda\textrm{CDM}$ cosmology with the Hubble constant value of $H_{0} = 70\,\mathrm{km\,s^{-1}\,Mpc^{-1}}$, $\Omega_{\mathrm{m}} = 0.3$, and $\Omega_{\Lambda} = 0.7$ \citep{2011ApJS..192...18K}.

\section{WALLABY \HI data}
\label{sec:data}
WALLABY is an extragalactic \HI survey being conducted with the Australian Square Kilometre Array Pathfinder (ASKAP; \citealt{2021PASA...38....9H}), a 36-antenna radio interferometer and precursor to the SKA. Equipped with innovative Phased Array Feeds (PAFs), ASKAP delivers a large instantaneous field of view ($\sim 30~\textrm{deg}^2$), making it an ideal instrument for wide-field \HI surveys. Upon completion, the full WALLABY survey will map approximately 14,000~$\textrm{deg}^2$ of the southern sky ($ -60^\circ \le \delta \le +10^\circ $) up to a redshift of $z \approx 0.1$. Prior to full survey operations, a series of pilot observations were conducted to assess data quality and optimise the observing and processing strategies for the main survey.

Throughout this paper, we use the publicly released \HI data from the WALLABY pilot survey. The sample comprises a total of 2,323 galaxies sourced from the first pilot data release (PDR1, \citealt{2022PASA...39...58W}) and the second pilot data release (PDR2, \citealt{2024PASA...41...88M}). These datasets consist of the Hydra, Norma, and NGC 4636 fields from PDR1, together with the NGC 4808, Vela, and NGC 5044 fields from PDR2.

\begin{figure}
\centering
\includegraphics[width=1\linewidth]{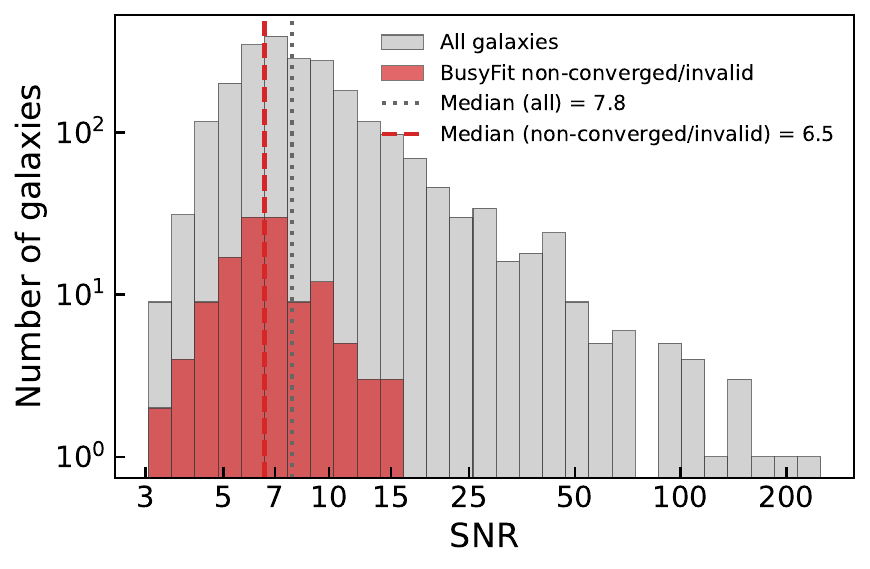}
\caption{Distribution of integrated signal-to-noise ratio (SNR) for the WALLABY pilot sample, displayed on a logarithmic scale. The grey and red histograms represent the full sample and galaxies where BusyFit failed (due to convergence failures or invalid linewidth measurements), respectively. The vertical dotted and dashed lines indicate the median SNR values for the full sample and this subsample, respectively, calculated in linear space.}
\label{fig:intSN_histogram}
\end{figure}

The WALLABY observations have a spectral resolution of $18.5~\mathrm{kHz}$, which translates into a velocity resolution of approximately $4~\mathrm{km \, s^{-1}}$ at zero redshift. The angular resolution is $30~\mathrm{arcsec}$, sufficient to at least marginally resolve the vast majority of galaxies detected by WALLABY \citep{2021MNRAS.506.3962W}. he root mean square (rms) noise level in the WALLABY data cubes is typically $\sim1.6~\mathrm{mJy\,beam^{-1}}$ at the original spatial and spectral resolution, but varies significantly as a function of both frequency and position on the sky, with the noise increasing markedly towards the edges of each pilot survey field. Figure~\ref{fig:intSN_histogram} shows the distribution of integrated signal-to-noise ratios (SNR) for the WALLABY pilot sample (grey histogram), a key parameter in our analysis. The SNR is defined as the ratio of the integrated flux to its statistical uncertainty \citep{2021MNRAS.506.3962W}:
\begin{equation}
\mathrm{SNR}
=
\frac{f_{\rm sum}}
{\sigma_{\rm rms}\,\Delta\nu
\sqrt{N_{\rm pix}/\Omega}},
\label{eg:SNR}
\end{equation}
where $f_{\textrm{sum}}$ is the integrated flux, $\sigma_{\textrm{rms}}$ is the local rms noise provided for each source in the WALLABY catalogue, $\Delta \nu$ is the channel width, $N_{\textrm{pix}}$ is the total number of pixels summed over all channels within the source mask, and $\Omega$ is the area of the Gaussian synthesised beam in pixel units. Both $\sigma_{\rm rms}$ and $N_{\rm pix}$ are provided for each galaxy in the WALLABY source catalogue, with the rms noise value assumed to be constant across all spectral channels for a given source. The distribution has a median SNR of 7.8, as indicated by the vertical dotted line in Figure~\ref{fig:intSN_histogram}. Since the observing strategy and sensitivity of the pilot survey are broadly representative of the full WALLABY survey, this distribution may provide an indication of the SNR distribution expected for the full survey. However, factors such as large-scale structure, artefacts, and any adjustments to source-finding parameters may alter this distribution in the full survey.

The WALLABY data release includes spatially-integrated spectra (global profiles) for each source. These are generated by the SoFiA source-finder after application of the SoFiA-generated mask. A consequence of SoFiA's masking approach is that the spectra do not contain signal outside of the frequency range covered by the source and hence no spectral ``baseline''. This creates a few minor problems for profile fitting which will be discussed later on.

\section{Methods}
\label{sec:methods}

We apply a spectral profile fitting technique to measure the linewidths from \HI global profiles of galaxies in a fully automated manner. This approach builds upon the Busy Function framework \citep{2014MNRAS.438.1176W}, which we have further developed to optimise performance for large-scale survey data. By fitting a physically motivated but versatile analytic function to the observed spectra, we mitigate the impact of local noise and provide a more robust estimation of the galaxy's rotational properties. The specific implementations and enhancements are detailed in the following subsections.

\subsection{Linewidth definitions}\label{sec:3.1}
Numerous linewidth definitions have been used in the literature. The following list summarises the various definitions which will be used in the following Sections, but is not a complete set:

\begin{enumerate}
    \item $\ensuremath{W_{20}}$: the linewidth measured at 20\% of the peak flux density of the integrated \HI spectral profile. This classical definition has been widely adopted in early Tully-Fisher studies \citep[e.g.][]{1977A&A....54..661T,1981ApJS...47..139F}. The 20\% level was empirically chosen as a compromise between sensitivity to the profile steepness, which is influenced by the detailed distribution of \HI within the galaxy, and sensitivity to noise in the observed spectrum.
    \item $\ensuremath{W_{50}}$: The linewidth measured at 50\% of the peak flux density of the integrated \HI spectral profile. Compared to $\ensuremath{W_{20}}$, the 50\% level is generally easier to determine in low signal-to-noise spectra. Consequently, $\ensuremath{W_{50}}$ has become one of the most commonly adopted linewidth definitions in large \HI surveys and catalogues \citep[e.g.][]{1990A&AS...82..391B,2004MNRAS.350.1195M,2011AJ....142..170H}. 
    \item  $\ensuremath{W_{m50}}$: The linewidth measured at 50\% of the mean flux density of the integrated \HI spectral profile, as used by \citet{2005ApJS..160..149S} and \citet{2009AJ....138.1938C}. Here, the mean is taken over the spectral range containing the inner 90\% of the integrated flux. This definition was introduced to reduce the dependence on peak flux measurements and to minimise the effects of noise, profile asymmetry, and variations in profile steepness. 
    \item  $\ensuremath{W_{F20}}$: The linewidth measured at 20\% of the nearest peak in the spectral profile \citep{1987MNRAS.224..953S, 2005ApJS..160..149S, 2018ApJ...861...49H}. Unlike $\ensuremath{W_{20}}$, $\ensuremath{W_{F20}}$ is typically derived from fits to the profile edges rather than from direct threshold crossings, reducing sensitivity to profile asymmetries and spectral noise.
    \item  $\ensuremath{W_{F50}}$: The linewidth measured at 50\% of the nearest peak in the spectral profile \citep{1987MNRAS.224..953S, 2005ApJS..160..149S, 2018ApJ...861...49H}. Similar to $\ensuremath{W_{F20}}$, this definition is typically derived from fitted profile edges and references each side of the profile to its nearest peak.
    \item  $\ensuremath{W_{75}}$: The linewidth measured at 75\% of the integrated flux constructed from the curve-of-growth (CoG) of the \HI spectral profile. Curve-of-growth-based linewidths were originally introduced by \citet{2020ApJ...898..102Y} as a way to parameterise the integrated \HI spectral profile in a robust framework based on the cumulative flux distribution, and were further developed by \citet{2023ApJ...950...87B}. In this paper, we extend the CoG linewidth definition from \citet{2023ApJ...950...87B} to additional flux fractions of 80\%, 85\%, and 90\%, denoted as $\ensuremath{W_{80}}$, $\ensuremath{W_{85}}$, and $\ensuremath{W_{90}}$, respectively. See Section~\ref{subsec:cog} for more details.
\end{enumerate}

In general, global \HI profiles will be noisy, so implementation of the above definitions may include masking, smoothing, interpolation, fitting, or application of width-minimising/maximising techniques \citep{2004MNRAS.350.1195M}.
However, an advantage of spectral profile fitting methods, including BusyFit, is that the profile is automatically de-noised, so these implementation details become much less important.

\subsection{The Busy function}\label{sec:3.2}

The Busy function is an analytic model specifically designed to fit the integrated \HI spectra of galaxies \citep{2014MNRAS.438.1176W}. It consists of a combination of error functions and a polynomial, using up to eight free parameters to capture the diverse morphology of \HI profiles. These parameters include: the total amplitude scaling factor ($a$), the steepness of the profile flanks ($b_1$, $b_2$), the half-width of the profile ($w$), the centroid of the overall profile ($x_e$), the centroid of the polynomial trough ($x_p$), the amplitude scaling factor of the trough ($c$), and the polynomial order of the trough ($n$). The mathematical form is expressed as:
\begin{equation}
\begin{split}
B(x) &= \frac{a}{4} \times 
\left( \text{erf}\left[b_1 \left( w + x - x_e \right) \right] + 1 \right) \\
&\quad \times \left( \text{erf}\left[b_2 \left( w - x + x_e \right) \right] + 1 \right) \times \left( c \left| x - x_p \right|^n + 1 \right).
\end{split}
\label{eq:first}
\end{equation}

\begin{figure}
\centering
\includegraphics[width=\linewidth]{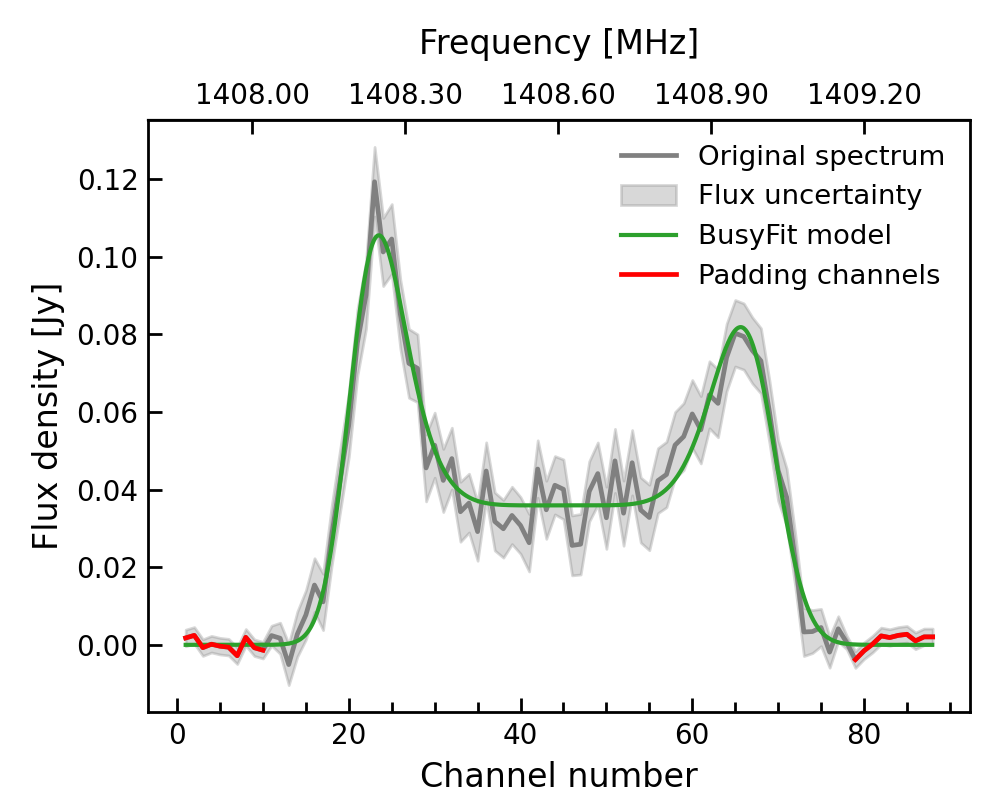}
\hspace{0.05\linewidth}
\caption{\label{fig:baseline}The \HI global spectrum of J101434-274133. The dark grey line is the original spectrum, while the red lines indicate an additional artificial padding of 10 spectral channels on either side. The light grey shaded region represents the flux uncertainty, which is sampled from a normal distribution with a standard deviation set to the mean flux uncertainty of the first and last unmasked spectral channels. The green line is the BusyFit model spectrum.}
\label{fig:Fig1}
\end{figure}

The Busy function is versatile and capable of accurately modeling a wide range of realistic profile shapes including Gaussian, double-horn, and flat-topped distributions  (see Figures~\ref{fig:Fig1} and \ref{fig:Fig2}). The primary advantage of extracting linewidths from Busy Function models instead of raw spectra is better robustness to noise, which is an issue that is largely unavoidable in direct parameterization methods such as those used by SoFiA. 

\subsection{Modifying and automating BusyFit}\label{sec:3.3}
The public BusyFit\footnote{Available at \url{https://gitlab.com/SoFiA-Admin/BusyFit}.} code \citep{2014MNRAS.438.1176W} uses $\chi^{2}$ minimisation to find the optimal parameters of the Busy function for a given spectral profile. However, its convergence reliability for large numbers of profiles is insufficient for our purposes, as BusyFit will typically struggle to converge on a reliable fit for low-SNR spectra, particularly at SNR values below 10. To address this limitation, we implement three procedures within our BusyFit-based pipeline that greatly improve fitting reliability: (i) incorporating flux density uncertainties; (ii) growing the spectral baseline; and (iii) providing informed initial parameter estimates.

Firstly, we calculate flux density uncertainties ($\Delta f_{\nu}$) from the local rms noise level in the vicinity of the source ($\sigma_{\rm rms}$) and the number of pixels ($N_{\nu}$) within the source mask in each frequency channel as follows:
\begin{equation}
\Delta f_{\nu} = \sigma_{\rm rms} \times \sqrt{\frac{N_{\nu}}{\Omega}} .
\end{equation}
For the WALLABY data used in this work, the restoring beam is \(30''\), corresponding to \(\Omega \approx 28.3\) pixels for a \(6''\) pixel size. When $N_{\nu} < \Omega$, we adopt $\Delta f_{\nu} = \sigma_{\rm rms}$ to avoid unrealistically small uncertainty estimates when the source occupies less than one beam area. Incorporation of these channel-dependent flux density uncertainties enables weighted fitting and the calculation of a meaningful reduced $\chi^2$, thereby improving the assessment of fit quality and convergence.

Secondly, for growing the spectral baseline, we append 10 artificial spectral channels to each edge of the SoFiA profile (see Figure~\ref{fig:baseline}). These channels are intended to provide a baseline region that stabilises the BusyFit fitting procedure and better constrains the profile as it approaches zero flux density outside the detected emission. The choice of 10 channels represents a practical balance: smaller padding lengths may not provide sufficient baseline to constrain the fit beyond the source profile, whereas substantially larger padding lengths can cause the baseline region to dominate the fit statistics. In the WALLABY pilot data, this padding length corresponds to a baseline extension of roughly 20\% of the profile width on each side, given the channel width of 18.5 kHz and the average SoFiA $\ensuremath{W_{20}}$ linewidth of approximately 880 kHz. As with any spectral profile fitting procedure that includes an excessive amount of baseline, this can artificially drive the reduced $\chi^{2}$ towards unity even when the source itself is poorly modelled. The flux densities assigned to these padding channels are drawn from a Gaussian distribution with zero mean and an rms based on the average uncertainty of the first and last valid channels in the spectrum.

Finally, to provide informed initial guesses for the fitting parameters, we derive initial estimates for the total amplitude scaling factor ($a$), centroid positions ($x_e$, $x_p$), and half-width ($w$) from closely related quantities provided by the SoFiA catalogue. Specifically, we use the maximum flux density ($f_{\rm max}$), the flux-weighted barycentric frequency ($\nu_{\rm centre}$), and the linewidth measured at 20\% of the peak flux density ($W_{20}$) to construct these initial guesses. We adopt the approximate relations between the SoFiA parameters and initial BusyFit parameters as given in Table~\ref{tab:initpars}.

\begin{table}
\centering
\caption{Initial estimates supplied to BusyFit based on WALLABY source catalog parameters.}
\label{tab:initpars}
\begin{tabular}{ll}
\hline
BusyFit   & Initial \\
parameter & estimate \\
\hline
$a$         & $f_{\rm max} / 2$ \\
$b_1$       & $0.5$ \\
$b_2$       & $0.5$ \\
$c$         & $0.0$ \\
$n$         & $2$ \\
$w$         & $W_{20} / 2$ \\
$x_{\rm e}$ & $\nu_{\rm centre}$ \\
$x_{\rm p}$ & $\nu_{\rm centre}$ \\
\hline
\end{tabular}
\end{table}

We evaluated the performance of four different fitting methods based on two primary criteria: the numerical convergence rate and the physical reliability of the resulting profiles. The former is determined by the BusyFit success flag, automatically reported with the fit results, where a value of 0 indicates a fully successful fit, 1 a converged solution with some parameters negative, and 2 a convergence failure.  Fits with success flags of 0 or 1 are considered numerically converged; however, cases with invalid linewidth measurements (i.e., negative or non-finite linewidths) are additionally excluded. Therefore, we define the success rate as the fraction of fits that both satisfy a BusyFit success flag of 0 or 1 and have valid linewidth measurements. However, since a numerically converged solution does not necessarily guarantee an accurate representation of the observed spectrum, we conducted a visual inspection of the fitted profiles. Consequently, the methods are ordered below from most to least reliable, reflecting a prioritised preference for models that accurately capture the physical properties of the observed \HI global spectra rather than mere mathematical convergence:

\begin{itemize}
\item  Method A: baseline padding included; initial parameters derived from SoFiA catalogue parameters. This method achieved a convergence success rate of 94.7\% (2,199/2,323). 
\item  Method B: no baseline padding; initial parameters derived from SoFiA catalogue parameters. This method resulted in a convergence success rate of 91.2\% (2,118/2,323).
\item  Method C:  baseline padding included; no pre-defined initial parameters. This method achieved a convergence success rate of 96.6\% (2,244/2,323).
\item  Method D: no baseline padding; no pre-defined initial parameters. This method achieved a convergence success rate of 96.4\% (2,240/2,323).
\end{itemize}

\begin{figure*}
\centering
    \includegraphics[width=0.49\linewidth]{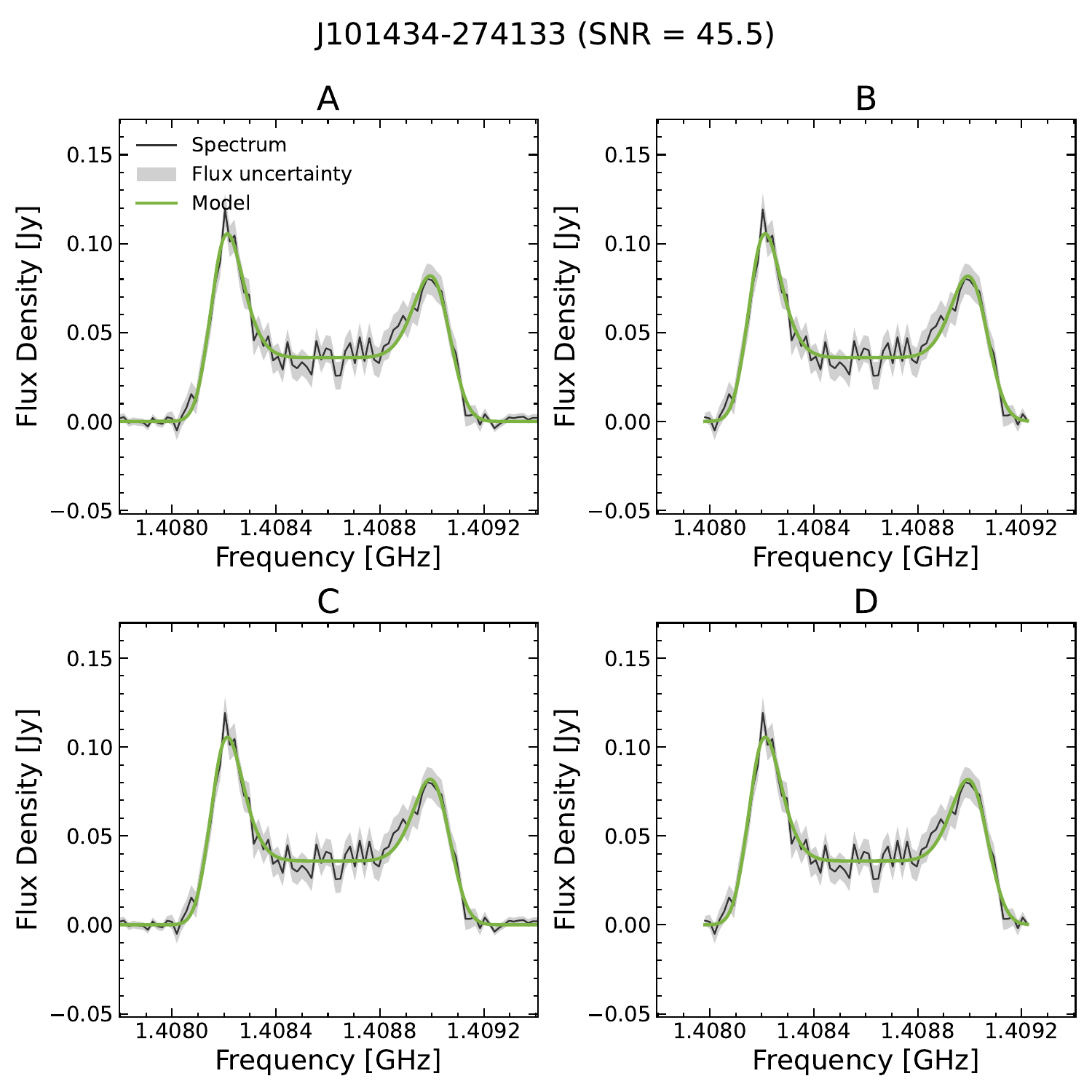}
    \includegraphics[width=0.49\linewidth]{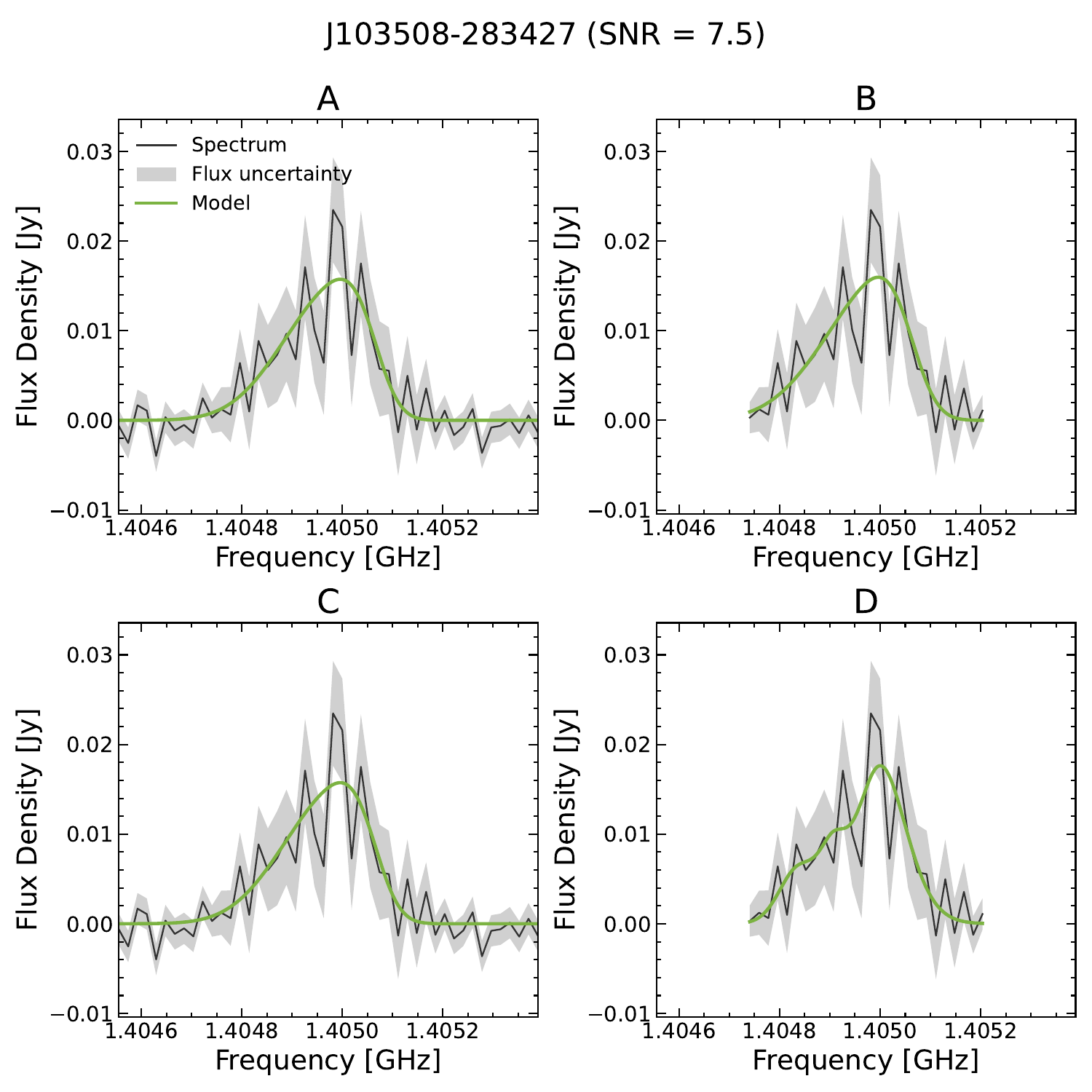}
    \caption{Comparison of BusyFit methods for the \HI global spectra of two galaxies: J101434-274133 (left; high SNR; double-peak profile) and J103508-283427 (right; low SNR; single-peak profile). For each panel: method A is with extra baseline and initial parameters; method B is without extra baseline but with initial parameters; method C is with extra baseline but without initial parameters; method D is without extra baseline and without initial parameters.}
    \label{fig:Fig2}
\end{figure*}

Figure~\ref{fig:Fig2} illustrates the ideal performance of the four BusyFit approaches (methods A--D) applied to the \HI global spectra of two example WALLABY galaxies: a high SNR double-horn profile (J101434-274133) and a low SNR single-peaked profile (J103508-283427). In these optimal cases, all four methods yield successful fits based on both numerical convergence and visual inspection, showing no significant differences between them.

\begin{figure*}
    \centering
    \includegraphics[width=0.49\linewidth]{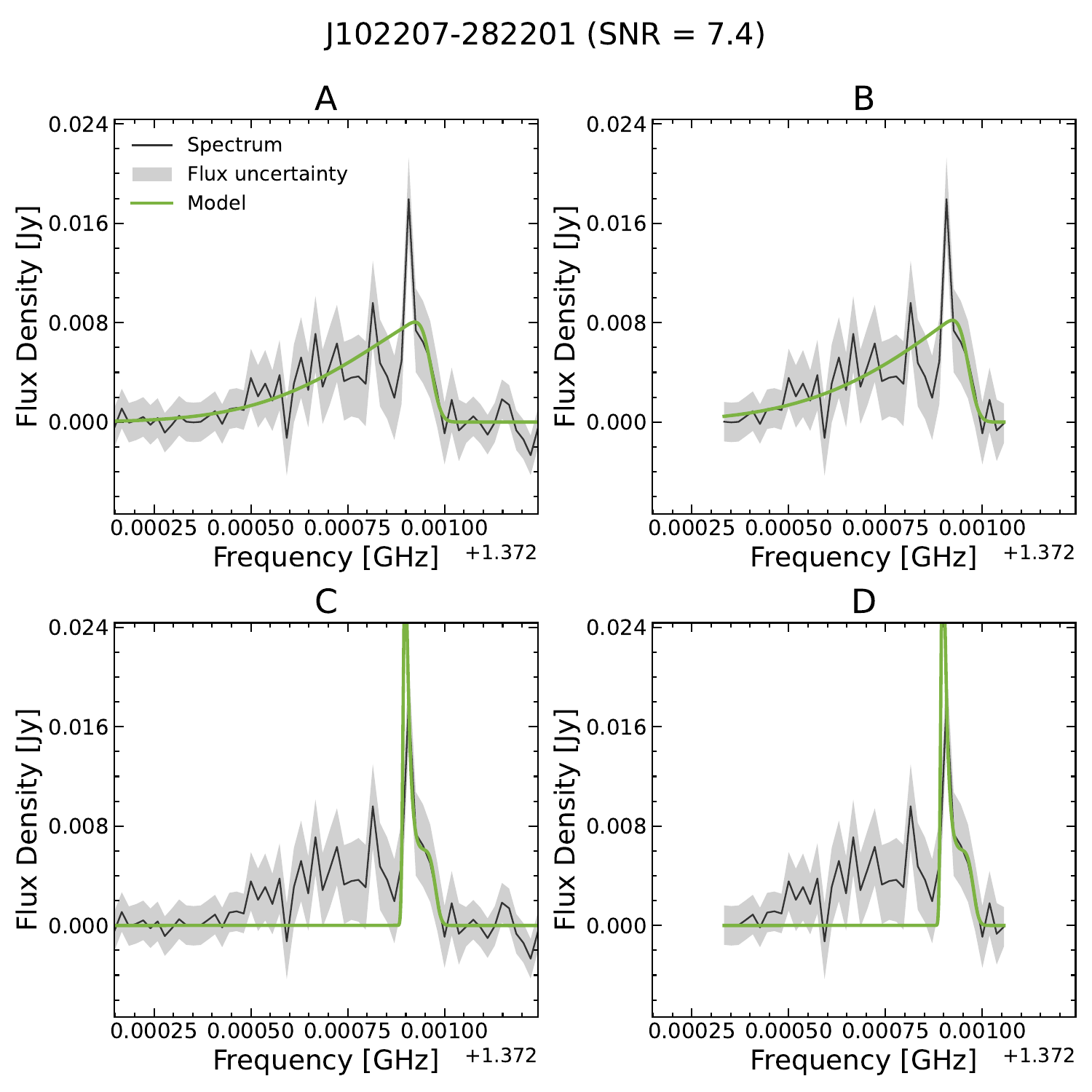}
    \includegraphics[width=0.49\linewidth]{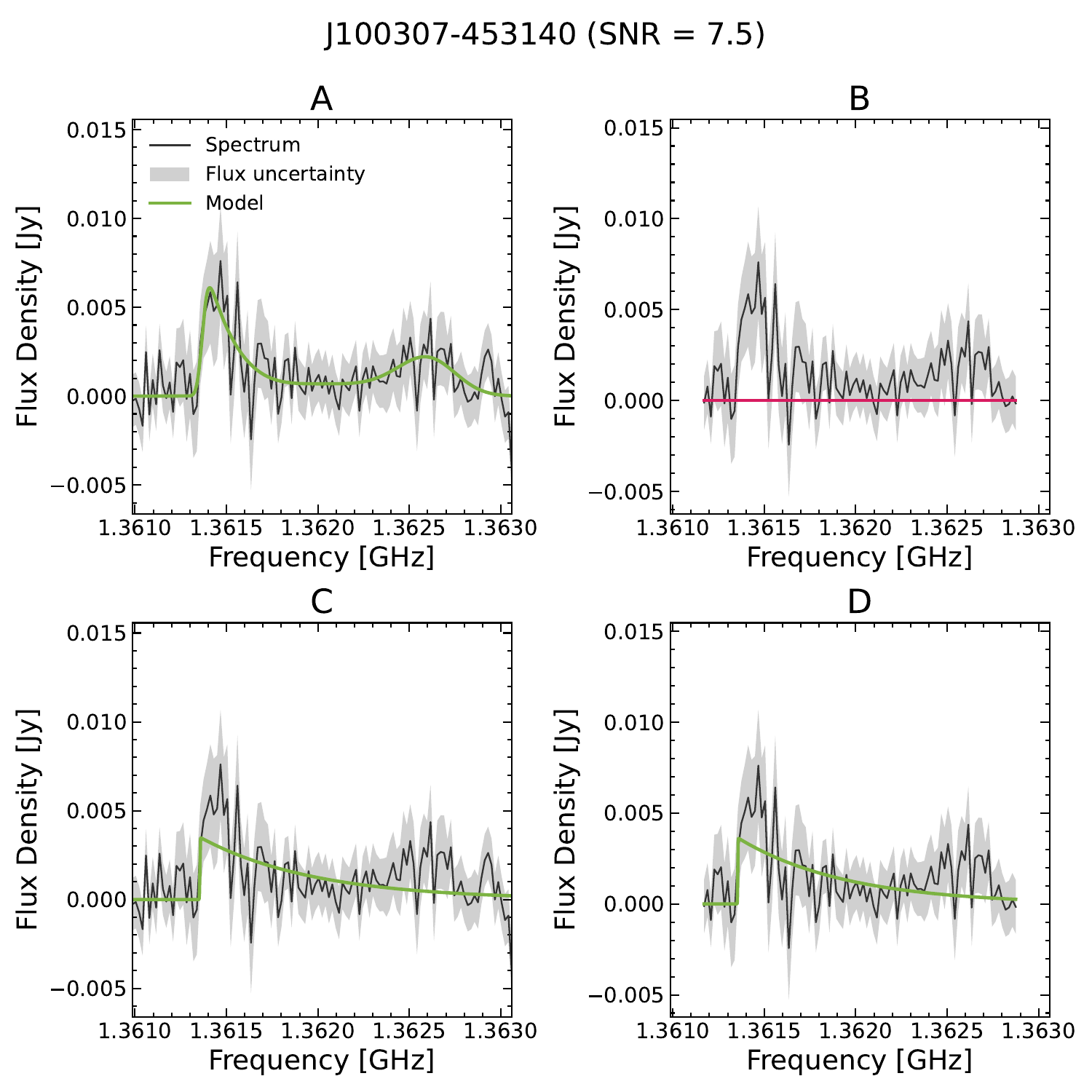}
    \caption{Comparison of BusyFit methods for the \HI global spectra of two galaxies: J102207-282201 (left) and J100307-453140 (right). The fitting methods are the same as those shown in Figure~\ref{fig:Fig2}. In the right panel, the method B model fit is shown in red to indicate that BusyFit failed to converge.}
    \label{fig:Fig3}
\end{figure*}

However, formal numerical convergence does not always imply a physically meaningful solution. Figure~\ref{fig:Fig3} illustrates two examples that highlight different failure modes. In the left-hand panels (J102207-282201), methods C and D, which do not use SoFiA-derived initial parameter estimates, have been trapped in a local minimum, whereas methods A and B successfully converge to the global minimum. In the right-hand panels (J100307-453140), method B fails to converge despite using the same SoFiA-derived initial parameter estimates as method A, while methods C and D formally converge but reproduce only part of the \HI emission profile. Only method A, which combines baseline padding with informed initial parameter estimates, converges to the physically meaningful solution.

Consequently, although methods C and D show marginally higher raw numerical convergence rates, visual inspection of the fitted profiles reveals that a fraction of these numerically converged solutions do not provide physically reliable representations of the observed profiles. In particular, these include cases where the BusyFit model captures only one side of a double-horned profile rather than the full global profile, resulting in poor fits with large reduced $\chi^{2}$ values. Such solutions contribute to the extended tails in the reduced $\chi^{2}$ distributions of methods C and D. Defining the extended tail as reduced $\chi^{2}$ > 1.2, approximately the 99th percentile of method A, fits in this tail occur about four times more frequently for methods C and D than for method A. In contrast, method B has the lowest numerical convergence rate, consistent with the failed example shown in Figure~\ref{fig:Fig3}, where BusyFit does not converge without baseline padding despite using appropriate initial parameter estimates. Taken together, these examples show that both baseline padding and appropriate initial parameter constraints are required to achieve more robust and physically meaningful fits and reliable linewidth measurements.

We therefore adopt method A for all subsequent analyses in this study, as it provides the highest rate of physically meaningful convergence. Despite its robustness, method A still fails to converge for a small fraction (5.3\%) of the sample, primarily due to limited SNR. The SNR distribution of these failed cases is shown in Figure~\ref{fig:intSN_histogram} as a red histogram. The failed cases are preferentially found at lower SNR, with a median SNR of 6.5 compared with 7.8 for the full sample. Furthermore, some of these failures involve partially detected galaxies in \HI, close galaxy pairs, or sources with irregular profiles that deviate significantly from standard \HI profiles.

\begin{figure*}
\centering
\includegraphics[width=1.0\linewidth]{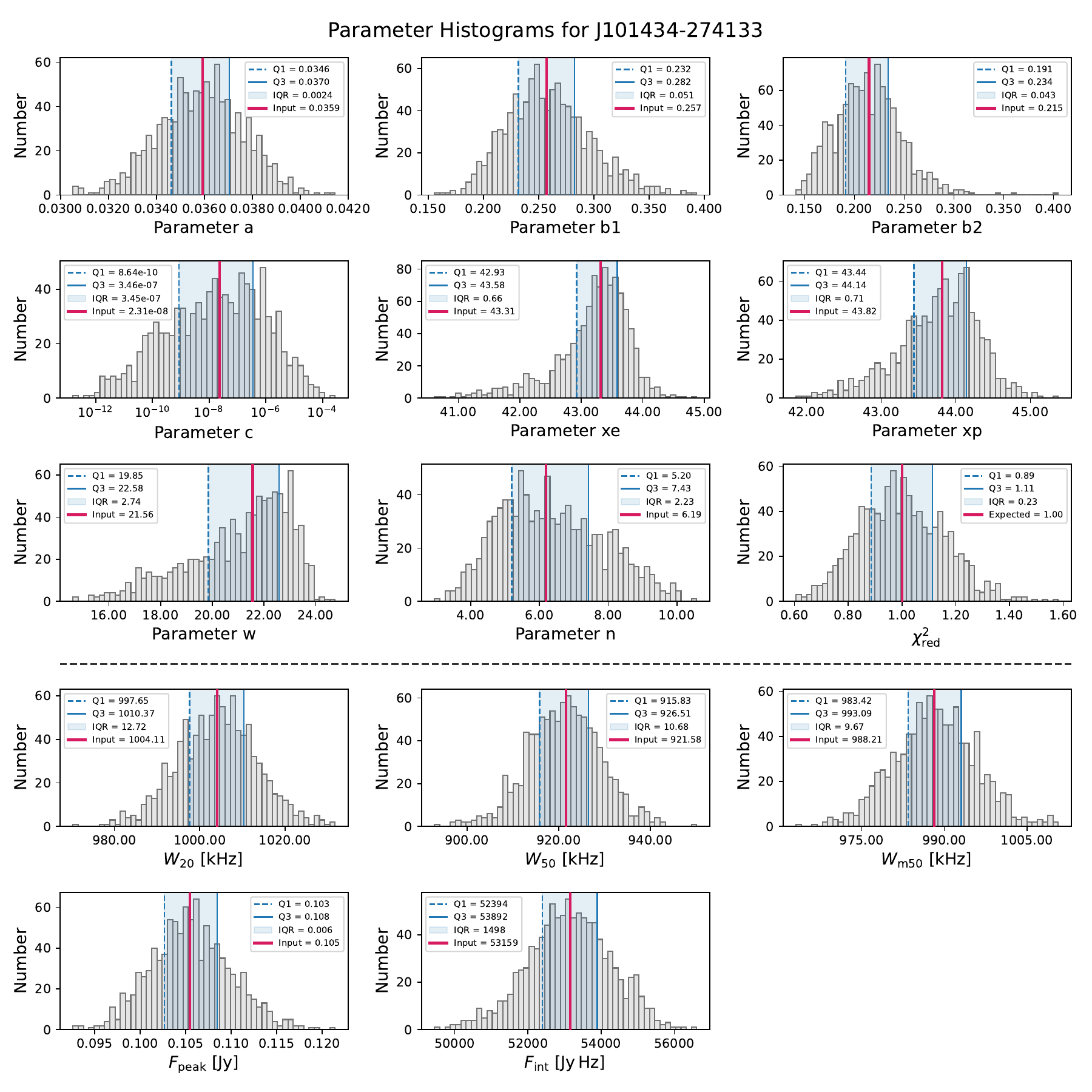}
\caption{\label{fig:Paramter_Distribution}Distribution of BusyFit parameters and derived parameters obtained from fitting the BusyFit to 1,000 mock integrated spectra of J101434-274133, the galaxy shown in Figure~\ref{fig:Fig2}. The top nine histograms (above the dashed separator line) show the BusyFit model parameters; the bottom six histograms refer to the derived parameters calculated from the BusyFit model. The vertical red lines indicate the input model parameters, while for  $\chi^{2}_{\mathrm{red}}$, the expected value of 1 is shown. The blue-shaded regions indicate the interquartile range (IQR). For parameter c, the histogram is displayed using logarithmically spaced bins on a logarithmic x-axis.}
\end{figure*}

\subsection{BusyFit linewidth uncertainties}\label{sec:BusyFiterrors}

The original BusyFit framework offers two internal methods for uncertainty estimation: (1) linear error propagation using the covariance matrix obtained from the $\chi^{2}$ minimisation; and (2) parameter variation, where the parameters are perturbed away from their best-fit values to assess the effect on the derived linewidth parameters. As will be shown later, both methods result in biased linewidth errors.

To address this problem, we propose a more robust method using Monte Carlo simulations. First, we generate 1,000 mock global spectra by adding Gaussian random noise to the best-fit BusyFit model. Here, we assume that the model provides an adequate description of the underlying spectral profile. The standard deviation of the Gaussian noise is determined by the per-channel flux uncertainties. Secondly, we re-apply the BusyFit methodology of Section~\ref{sec:3.3} to each of these 1,000 mock spectra to build a distribution of parameter estimates. Thirdly, we quantify the parameter uncertainties using the interquartile range (IQR), defined as the difference between the 25th and 75th percentiles of the distributions obtained from the 1,000 Monte Carlo realisations.

As illustrated in Figure~\ref{fig:Paramter_Distribution}, which is an example for one galaxy profile (J101434-274133), most parameters follow an almost Gaussian distribution, but with extended tails. The red solid line represents the reference value, which is the BusyFit best-fit model parameter. In all cases, the parameter distributions are centered close to this reference value (i.e.\ low parameter bias). Due to the non-Gaussian tails, it is more appropriate to quantify the uncertainties using the interquartile range (IQR) for each parameter distribution. This approach ensures more reliable parameter uncertainty estimates compared to the methods implemented within BusyFit.

\begin{figure*}
\centering
\includegraphics[width=1.0\linewidth]{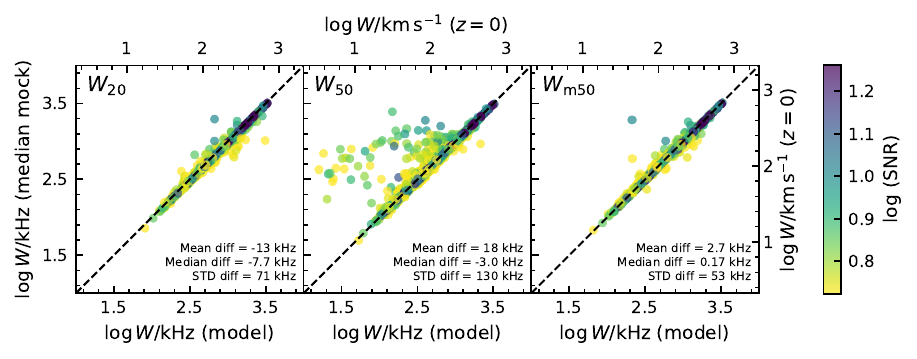}
\caption{Comparison between the model linewidths (derived from BusyFit fits to 2,199 WALLABY pilot survey profiles using method A) and the median values derived from BusyFits to 1,000 mock noise spectra added to each of the models. The panels show BusyFit linewidths $\ensuremath{W_{20}}$, $\ensuremath{W_{50}}$, and $W_{\mathrm{m50}}$. Statistical differences are indicated in the bottom-right corner of each plot on a linear scale. All widths are shown in kHz on logarithmic axes, with the dashed line representing the one-to-one relation. The top and right-hand secondary axes show the corresponding
linewidths in velocity units of $\mathrm{km\,s^{-1}}$ on logarithmic scales,
calculated at $z=0$ using the \HI rest frequency.} The points are colour-coded by logarithmic SNR.
\label{fig:6}
\end{figure*}

Figure~\ref{fig:6} compares the input model linewidths obtained from the 2,199 BusyFit models obtained using method A with the median linewidths recovered from 1,000 mock spectra generated for each model. The points are colour-coded by logarithmic SNR. Overall, our method successfully reproduces the input parameters, demonstrating that the fitting pipeline successfully recovers the true underlying values without significant global systematics. The model values for all of these parameters are generally consistent with the median mock values. However, $\ensuremath{W_{50}}$ shows a notable number of outliers where the median values for the mock spectra are larger than the input BusyFit model. This discrepancy is primarily observed at SNR below 10, where noise significantly affects the reliability of the fit. In these cases, the BusyFit model can occasionally follow noise-induced peaks near the profile maximum, producing an artificially peaked profile and hence a narrower $\ensuremath{W_{50}}$. When the model is subsequently refitted to many independent noise realisations, these narrow features are not recovered consistently, and the fits tend to converge on slightly broader profiles, resulting in larger median $\ensuremath{W_{50}}$. The only notable outlier in $\ensuremath{W_{m50}}$ occurs despite its high SNR and is associated with a case where BusyFit converges numerically but fails to provide a physically reliable representation of the observed profile. For each panel in Figure~\ref{fig:6}, statistical metrics (the mean, median, and standard deviation of the differences) are provided in the bottom-right corner. Among the three linewidth definitions, $\ensuremath{W_{m50}}$ exhibits the smallest systematic offset and scatter, indicating the most consistent recovery of the input model linewidths.

\subsection{SoFiA linewidths}\label{sec:sofia}
The SoFiA 2 source-finding application \citep{2021MNRAS.506.3962W} also provides measurements for $\ensuremath{W_{20}}$, $\ensuremath{W_{50}}$, and $\ensuremath{W_{m50}}$. These are not from a fitting procedure, but from a width-maximising estimate directly from the spectral profiles. These provide useful quick measurements and are available in the released WALLABY source catalogs \citep{2022PASA...39...58W, 2024PASA...41...88M}. However, some of them may be subject to biases when applied to spectra of low SNR because the peak flux density for the $\ensuremath{W_{20}}$ and $\ensuremath{W_{50}}$ measurements is obtained directly from the spectrum and therefore biased by noise (see \citealt{2021MNRAS.506.3962W} for details).

\begin{figure*}
\centering
\includegraphics[width=0.7\linewidth]{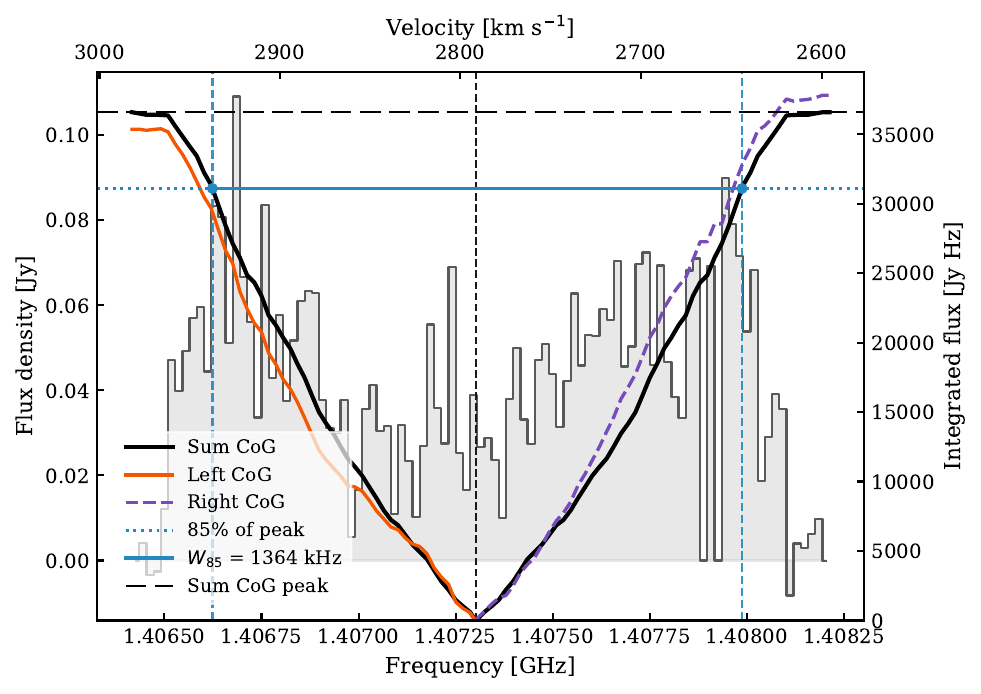}
\caption{Illustration of the curve-of-growth (CoG) method applied to the BusyFit model of J100351-273417. The bottom x-axis shows frequency (GHz), while the top x-axis shows systemic velocity ($\mathrm{km\,s^{-1}}$). Starting from the central frequency (black vertical dashed line), the cumulative flux is summed outward to the left and right, shown in orange and purple, respectively. The sum CoG (black solid line) is defined as the average of the two curves. The linewidth $\ensuremath{W_{85}}$ is measured at 85\% of the peak value of the sum CoG. The corresponding frequency boundaries are indicated by the vertical blue dashed lines, and the 85\% level is marked by the blue dotted horizontal line. The resulting linewidth $\ensuremath{W_{85}}$ is highlighted by the blue solid line.}
\label{fig:CoG}
\end{figure*}

\subsection{Curve-of-growth (CoG) linewidths}
\label{subsec:cog}
The curve-of-growth (CoG) linewidth estimator for the \HI profiles was introduced by \citet{2020ApJ...898..102Y} and \citet{2023ApJ...950...87B} to fit large numbers of profiles in a robust fashion. The CoG is generated by integrating the flux density outward from the profile centre toward the edges and averaging the resulting curves on both sides. This approach was originally designed to provide robust, automated linewidth measurements for ALFALFA survey galaxies, specifically by treating the outer 15\% or 25\% of the spectral edges as noise-dominated regions. The primary advantage of the CoG method is its reliance on integrated flux rather than local extrema (e.g.\ peak flux density or edge steepness). Consequently, it may be more resilient to noise fluctuations in low SNR spectra compared to $\ensuremath{W_{20}}$ or $\ensuremath{W_{50}}$. Our implementation of the CoG algorithm is more streamlined, mainly reflecting the fact that the global profiles produced by SoFiA are generally masked, which automatically defines the spectral extent of the profile. 
Our modified procedure is illustrated in Figure~\ref{fig:CoG}, and summarised as follows:
\begin{enumerate}
\item Centering: the flux-weighted barycentric frequency of the profile is determined from the spectrum on a per-source basis. This defines the centre of the profile, represented by the black vertical dashed line in Figure~\ref{fig:CoG}.
\item Integration: moving outward from this centre, we cumulatively sum the flux in subsequent channels to both the left (orange solid curve) and right (purple dashed curve). The summation is performed on the native spectral grid, starting from the channel closest to the adopted centre frequency. 
\item Averaging: the two integration curves are averaged to produce the final CoG (black solid curve).  
\item Peak identification: the maximum integrated flux of the final CoG is identified (black long dashed line).
\item Width extraction: using this maximum value (black long dashed line) as the 100\% flux level, we measure the velocity width at a specified percentage level (defined by the distance between two blue dots in Figure~\ref{fig:CoG}, here at 85\% of the maximum, $\ensuremath{W_{85}}$). The intersection points are obtained by linear interpolation between adjacent channels that bracket the target flux level.
\end{enumerate}

\section{Results}
\label{sec:results}

In this section, we compare the linewidth and uncertainty measurements obtained from our BusyFit implementation with those from other methods. 

\subsection{Comparison with SoFiA measurements}\label{sec:comparison}

\begin{figure*}
\centering
\includegraphics[width=1\linewidth]{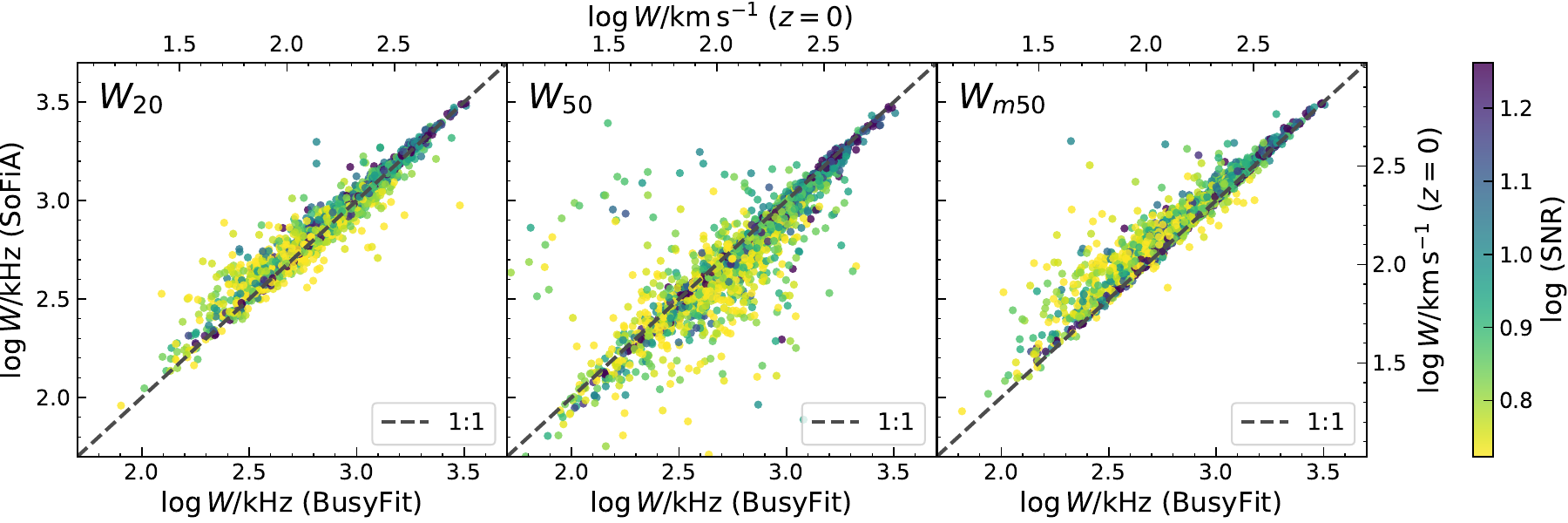}
\caption{\label{fig:Compare_SoFiA}Comparison of the logarithm of $\ensuremath{W_{50}}$, $\ensuremath{W_{m50}}$, and $\ensuremath{W_{20}}$ linewidths derived using BusyFit with the corresponding measurements from SoFiA. Linewidths are shown in kHz on logarithmic axes. The top and right-hand secondary axes show the corresponding linewidths in velocity units of $\mathrm{km\,s^{-1}}$ on logarithmic scales, calculated at $z=0$ using the \HI rest frequency.} The mean, median and rms differences are listed in Table~\ref{tab:line_width_statistics}. The colour bar indicates the SNR on a logarithmic scale, and the grey dashed line represents the one-to-one relation. 
\end{figure*}

The SoFiA 2 source finding pipeline \citep{2021MNRAS.506.3962W} provides linewidth measurements of $\ensuremath{W_{20}}$, $\ensuremath{W_{50}}$, and $\ensuremath{W_{m50}}$. The measurement comparisons are shown in Figure~\ref{fig:Compare_SoFiA}, where the points are colour-coded by logarithmic SNR. The middle panel reveals that $\ensuremath{W_{50}}$ measurements from SoFiA are lower on average than $\ensuremath{W_{50}}$ measurements derived using BusyFit. This discrepancy arises because SoFiA tends to underestimate the $\ensuremath{W_{50}}$ linewidth due to the impact of noisy flux values. Specifically, the peak flux density may be overestimated due to noise, or one of the peaks may be below 50\% of the other peak due to noise or intrinsic asymmetry. However, there are also a couple of dozen extreme outliers for which the SoFiA $\ensuremath{W_{50}}$ values are greater. These cases primarily occur in low SNR spectra (SNR < 10) where noisy features at both spectral edges lead to a discrepancy. While the BusyFit does not capture one of the peaks, leading to an underestimation of the width, SoFiA's broader integration encompasses these noisy edges, resulting in a larger measurement.

In contrast, the $\ensuremath{W_{20}}$ measurements from Busyfit and SoFiA are generally consistent, showing the tightest relation with the smallest mean and standard deviation of the differences. The $\ensuremath{W_{m50}}$ measurements from Busyfit and SoFiA are also generally consistent, although $\ensuremath{W_{m50}}$ values from SoFiA are slightly higher than those from BusyFit. A summary of the statistical differences between the two measurement methods is given in Table~\ref{tab:line_width_statistics}. This analysis underscores the effectiveness of deriving parameters through model fitting with BusyFit applied to integrated spectra.

\begin{table*}
\centering
\caption{Mean, median, and rms differences between linewidth measurements. The upper section corresponds to the BusyFit–SoFiA comparison in Figure~\ref{fig:Compare_SoFiA}, while the lower section shows the BusyFit–CoG comparison derived from observed WALLABY spectra in Figure~\ref{fig:CoG_comparison}. Velocity equivalents are calculated at $z=0$ using the \HI rest frequency.}
\label{tab:line_width_statistics}
\renewcommand{\arraystretch}{1.15}
\begin{tabular}{@{}llrrr@{\hspace{1.5em}}rrr@{}}
\toprule
\multirow{2}{*}{Method comparison}
 & \multirow{2}{*}{Linewidth}
 & \multicolumn{3}{c}{[kHz]}
 & \multicolumn{3}{c}{[km\,s$^{-1}$]} \\
\cmidrule(lr){3-5}\cmidrule(lr){6-8}
 & & Mean & Median & rms
 & Mean & Median & rms \\
\midrule

\multirow{3}{*}{BusyFit$-$SoFiA}
 & $W_{20}$
 & 63 & 36 & 94
 & 13.2 & 7.7 & 19.8 \\

 & $W_{50}$
 & 111 & 57 & 167
 & 23.4 & 11.9 & 35.3 \\

 & $W_{m50}$
 & 86 & 58 & 108
 & 18.2 & 12.3 & 22.8 \\

\midrule

\multirow{4}{*}{BusyFit$-$CoG}
 & $W_{m50}$ vs $W_{75}$
 & 182 & 163 & 117
 & 38.4 & 34.4 & 24.7 \\

 & $W_{m50}$ vs $W_{80}$
 & 142 & 125 & \change{103}
 & \change{30.1} & \change{26.5} & \change{21.7} \\

 & $W_{m50}$ vs $W_{85}$
 & \change{103} & \change{86} & \change{89}
 & \change{21.7} & \change{18.1} & \change{18.8} \\

 & $W_{m50}$ vs $W_{90}$
 & \change{69} & \change{54} & \change{75}
 & \change{14.5} & \change{11.3} & \change{15.9} \\

\bottomrule
\end{tabular}
\end{table*}

\subsection{Comparison with CoG measurements}\label{sec:cog_compare}
\begin{figure*}
\centering
\includegraphics[width=1\linewidth]{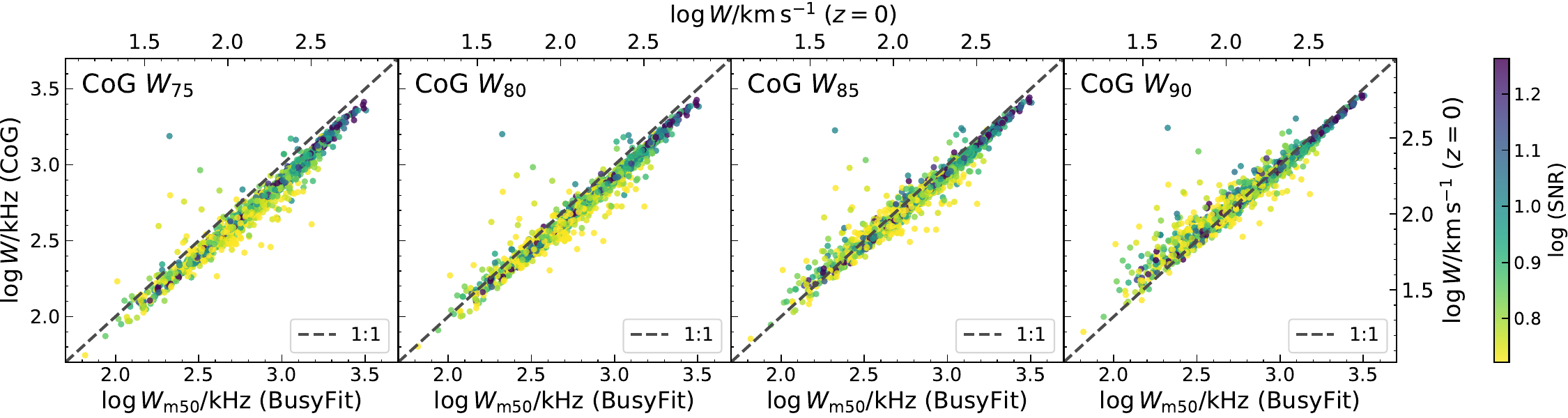}
\caption{Comparison of BusyFit $\ensuremath{W_{m50}}$ measurements with CoG linewidths derived from WALLABY spectra. The CoG linewidths are defined at 75\%, 80\%, 85\%, and 90\% of the total integrated flux, corresponding to $\ensuremath{W_{75}}$, $\ensuremath{W_{80}}$, $\ensuremath{W_{85}}$, and $\ensuremath{W_{90}}$, respectively. \change{Linewidths are shown in kHz on logarithmic axes. The top and right-hand secondary axes show the corresponding
linewidths in velocity units of $\mathrm{km\,s^{-1}}$ on logarithmic
scales, calculated at $z=0$ using the \HI rest frequency.} The mean, median, and rms differences are summarised in Table~\ref{tab:line_width_statistics}. The colour bar indicates the SNR on a logarithmic scale, and the grey dashed line represents the one-to-one relation. }
\label{fig:CoG_comparison}
\end{figure*}

We apply the CoG method to the observed WALLABY spectra as described in Section~\ref{subsec:cog}. We compute the CoG linewidths at 75\%, 80\%, 85\%, and 90\% of the total integrated flux, denoted as $\ensuremath{W_{75}}$, $\ensuremath{W_{80}}$, $\ensuremath{W_{85}}$, and $\ensuremath{W_{90}}$, respectively. In Figure~\ref{fig:CoG_comparison}, we compare the CoG linewidths with the BusyFit $\ensuremath{W_{m50}}$ measurements. The CoG $\ensuremath{W_{75}}$ and $\ensuremath{W_{80}}$ values are, on average, lower than the $\ensuremath{W_{m50}}$ linewidths, whereas the CoG $\ensuremath{W_{85}}$ and $\ensuremath{W_{90}}$ measurements show the closest agreement. Specifically, $\ensuremath{W_{85}}$ is generally consistent with the BusyFit $\ensuremath{W_{m50}}$ measurements but tends to be systematically lower in the large linewidth regime. This discrepancy arises because massive, fast-rotating galaxies typically exhibit steep spectral edges. This causes the cumulative flux in the CoG integration to rise sharply near the profile edges, shortening the apparent channel separation required to encompass a given fraction of the total flux and resulting in a narrower linewidth measurement. Despite this localised offset for wide profiles, $\ensuremath{W_{85}}$ remains the most consistent estimation for $\ensuremath{W_{m50}}$ overall. Since $\ensuremath{W_{m50}}$ is commonly used as a proxy for rotation velocity, this overall consistency suggests that $\ensuremath{W_{85}}$ is the most suitable CoG parameter for Tully-Fisher studies when the method is applied to masked profiles. Conversely, the $\ensuremath{W_{75}}$ parameter adopted in \citet{2023ApJ...950...87B} may be more appropriate for unmasked profiles, as it effectively excludes noise at the profile edges; however, this remains to be tested. The disadvantages of the CoG method include its reliance on a central velocity estimate and its inherent assumption of spectral symmetry. Nevertheless, the scatter is considerably lower than that in the BusyFit comparison with SoFiA linewidths, confirming the CoG measurements as a robust alternative.

\subsection{Uncertainty comparisons}\label{sec:error_compare}
As discussed in Section~\ref{sec:BusyFiterrors}, we have been able to generate robust error estimates for any spectrum able to be fit with a BusyFit model. This is achieved by adding noise spectra to the model and fitting a new BusyFit model 1000 times. This allows us to compare our error estimates with other estimates, where these exist, and also to compare the robustness of the different width definitions (see Section~\ref{sec:3.1}) and different algorithmic implementations (see Section~\ref{sec:comparison} and \ref{sec:cog_compare}).

\begin{figure*}
\centering
\includegraphics[width=1\linewidth]{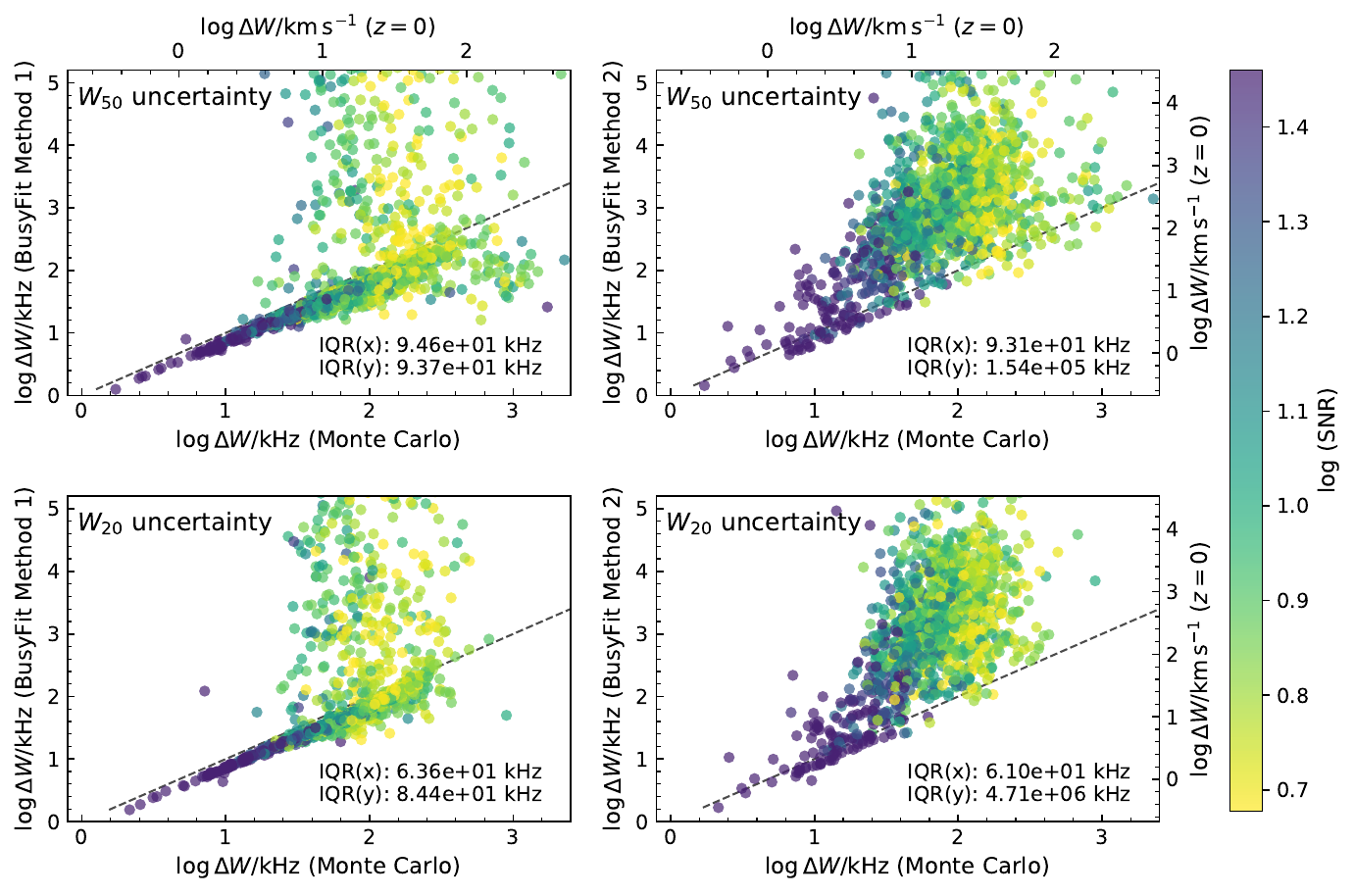}
\caption{\label{fig:Error_Compare}
Comparison of $\ensuremath{W_{50}}$ (top row)  and $\ensuremath{W_{20}}$ (bottom row) linewidth uncertainties measured using the two internal BusyFit methods (left and right columns, respectively) and the Monte Carlo method adopted in this analysis ($x$ axis). The Monte Carlo uncertainties are represented by the interquartile range (IQR) derived from the parameter distribution of 1000 mock spectra, as depicted in Figure~\ref{fig:Paramter_Distribution}. The grey dashed line represents the one-to-one relation, and the colour bar indicates the SNR on a logarithmic scale. \change{Linewidth uncertainties are shown in kHz on logarithmic axes. The top and right-hand secondary axes show the corresponding uncertainties in velocity units of $\mathrm{km\,s^{-1}}$ on logarithmic scales, calculated at $z=0$ using the \HI rest frequency. The inset IQR values are computed across the plotted sample for each axis in linear units of kHz.}}
\end{figure*}

Figure~\ref{fig:Error_Compare} compares our $\ensuremath{W_{20}}$ and $\ensuremath{W_{50}}$ linewidth uncertainties with the two internal BusyFit error estimates. The BusyFit method 1 (linear propagation of the covariance matrix) uncertainties are broadly consistent with our Monte Carlo approach. In particular, for high-SNR sources (SNR $>10$), the method 1 uncertainties generally agree well with the  Monte Carlo estimates, while being systematically smaller, with median ratios of $0.64$ and $0.70$ for $\ensuremath{W_{50}}$ and $\ensuremath{W_{20}}$, respectively. In contrast, the BusyFit method 2 (variation of parameters) uncertainties can be as large as $14$~dex higher, and we therefore consider method 2 not realistic for our sample. As shown by the color coding, the larger uncertainties typically arise in the low-SNR regime (SNR < 10), where the relative parameter uncertainties becoming large. While BusyFit method 1 seems to provide a reasonable uncertainty estimate, it is not straightforward to determine whether method 1 or the Monte Carlo is more accurate. However, assuming that the Busy Function provides an adequate representation of the observed spectrum, we adopt the Monte Carlo uncertainties as our fiducial linewidth uncertainties for the remainder of the analysis.

\begin{table*}
\centering
\caption{Best-fit parameters of the power-law model $\Delta W/W = A\,\mathrm{SNR}^{\alpha} W^{\beta}(W_{20}-W_{50})^{\gamma}$ derived for each linewidth definition. The linewidth ($W$), proxy for profile steepness ($W_{20}-W_{50}$), and the linewidth uncertainties ($\Delta W$) are all expressed in units of $\mathrm{km\,s^{-1}}$ in the rest-frame velocity domain. The final column, $\sigma_{\textrm{res}}$, gives the scatter of the residuals between the observed and modelled log $\Delta W$ values.}
\label{tab:3}

\begin{tabular}{l >{\centering\arraybackslash}m{1.8cm} c c c c c}
\hline
Method & Linewidth uncertainty & $A$ & $\alpha$ & $\beta$ & $\gamma$ & $\sigma_{\textrm{res}}$\\
\hline
\multirow{3}{*}{SoFiA}
 & $\Delta W_{20}/W_{20}$ & $14.58\pm1.15$ & $-0.59\pm0.01$ & $-0.79\pm0.02$ & $0.15\pm0.01$ & $0.16$ \\
 & $\Delta W_{50}/W_{50}$ & $4.13\pm0.28$ & $-0.93\pm0.01$ & $-0.58\pm0.01$ & $0.45\pm0.01$ & $0.15$ \\
 & $\Delta W_{m50}/W_{m50}$ & $26.64\pm2.43$ & $-0.61\pm0.02$ & $-0.91\pm0.02$ & $0.13\pm0.01$ & $0.18$ \\
\hline
\multirow{3}{*}{BusyFit}
 & $\Delta W_{20}/W_{20}$ & $5.48\pm0.19$ & $-1.04\pm0.01$ & $-0.55\pm0.01$ & $0.36\pm0.00$ & $0.07$ \\
 & $\Delta W_{50}/W_{50}$ & $7.16\pm0.32$ & $-1.25\pm0.01$ & $-0.54\pm0.01$ & $0.48\pm0.01$ & $0.10$ \\
 & $\Delta W_{m50}/W_{m50}$ & $5.88\pm0.21$ & $-1.12\pm0.01$ & $-0.58\pm0.01$ & $0.40\pm0.01$ & $0.08$ \\
\hline
\end{tabular}
\end{table*}

\begin{figure*}
\centering
\includegraphics[width=1\linewidth]{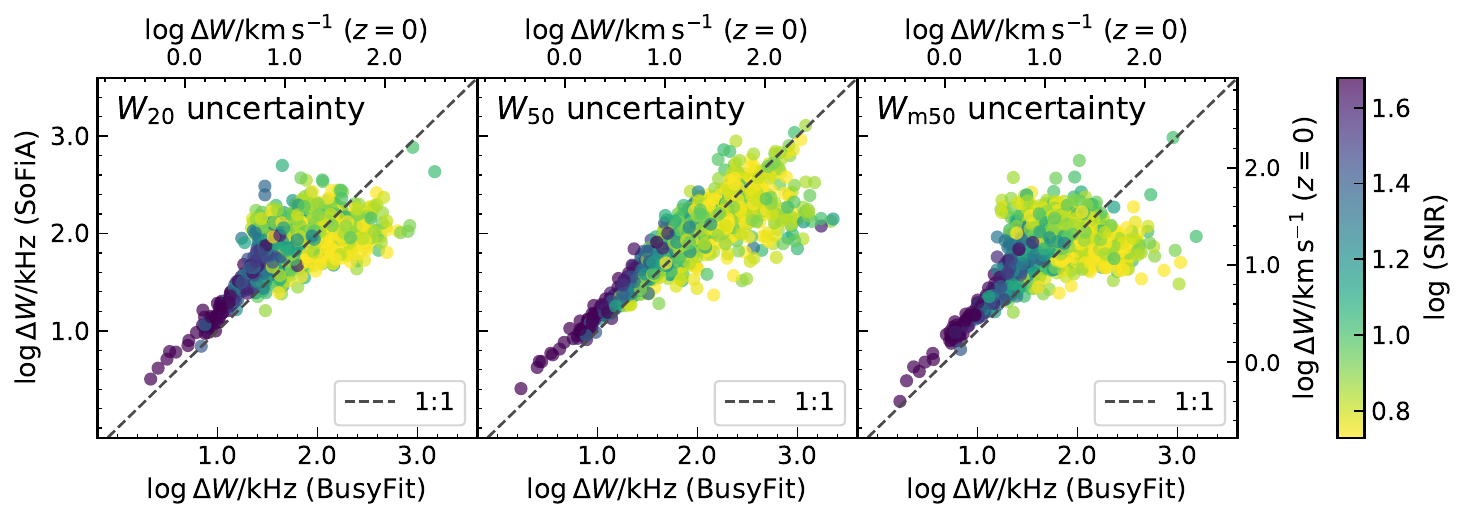}
\caption{Comparison of $\ensuremath{W_{20}}$, $\ensuremath{W_{50}}$, and $\ensuremath{W_{m50}}$ linewidth uncertainties measured using SoFiA and BusyFit algorithm. For each galaxy, 1,000 mock spectra generated from the BusyFit best-fit model are analysed independently by both algorithms. The uncertainty for each method is quantified using the interquartile range (IQR) of its respective linewidth distribution. \change{Linewidth uncertainties are shown in kHz on logarithmic axes. The top and right-hand secondary axes show the corresponding
linewidth uncertainties in velocity units of $\mathrm{km\,s^{-1}}$ on logarithmic
scales, calculated at $z=0$ using the H\,I rest frequency.} The colour bar indicates the SNR on a logarithmic scale, and the grey dashed line represents the one-to-one relation.}
\label{fig:IQR_SF_BF}
\end{figure*}

The effect of different algorithmic implementations has already manifested itself in the scatter in Figures~\ref{fig:Compare_SoFiA} and \ref{fig:CoG_comparison}, with the SoFiA approach having higher scatter than CoG measurements, when compared with BusyFit $\ensuremath{W_{m50}}$ measurements. However, our error estimates based on mock spectra more directly encode the uncertainties in each measurement. In Figure ~\ref{fig:IQR_SF_BF}, we compare the linewidth uncertainties obtained from SoFiA and BusyFit. Since SoFiA does not inherently calculate linewidth uncertainties, we derive them using the same Monte Carlo framework adopted for BusyFit. For SoFiA, however, the mock spectrum generation is modified to account for its use of masked spectral profiles. The original masked spectrum is first extended by adding 10 zero-flux channels at both edges, and BusyFit is fitted to this extended spectrum. After fitting, the artificially added channels are removed from the model spectrum, and Gaussian noise is injected to generate 1,000 mock realisations. The SoFiA linewidth algorithm is then applied independently to each mock spectrum, and the uncertainty is quantified using the IQR of the resulting linewidth distribution. This approach enables a direct comparison between SoFiA and BusyFit uncertainties because both are estimated from Monte Carlo simulations using the same observed spectra and equivalent noise levels. The resulting uncertainty distributions therefore capture the different sensitivities of the two algorithms to noise fluctuations in the spectral profile. As shown in Figure~\ref{fig:IQR_SF_BF}, the SoFiA uncertainties are generally larger than the BusyFit uncertainties for all linewidth definitions ($\ensuremath{W_{20}}$, $\ensuremath{W_{50}}$, and $\ensuremath{W_{m50}}$). This reflects the greater sensitivity of the SoFiA measurements to noise fluctuations, as the linewidths are determined directly from the observed spectral profile and are affected by noise-induced variations in the threshold crossings. In the low SNR regime, however, the SoFiA uncertainties become smaller than those of BusyFit. This behaviour is likely due to the different ways in which the two algorithms respond to noise-dominated spectra. In this regime, the linewidth measurement, especially $\ensuremath{W_{20}}$, becomes increasingly constrained by the extent of the source mask rather than by the underlying spectral profile. This can artificially stabilise the measured linewidth against noise fluctuations, resulting in underestimated uncertainties. However, this effect is accompanied by a systematic broadening of linewidth, as the mask can extend into noise-dominated regions. In contrast, BusyFit estimates the linewidth from the fitted model rather than from the extent of the source mask, and therefore reflects the increased uncertainty in determining the profile boundaries in noise-dominated spectra.

\subsubsection{Linewidth uncertainty parameterisation}
Following the studies of \citet{1986AJ.....92..742S, 1990A&AS...86..473F}, a number of large \HI surveys, including HIPASS \citep{2004AJ....128...16K}, ALFALFA \citep{2011AJ....142..170H, 2018ApJ...861...49H}, and FASHI \citep{2024SCPMA..6719511Z}, have commonly relied on empirical parameterisations to estimate linewidth uncertainties as a function of observational quantities such as the SNR and the steepness of the spectral profile edges. This approach provides a practical way to assign linewidth uncertainties to large galaxy samples without the need to perform computationally expensive Monte Carlo simulations for every source.

Although we derive our fiducial uncertainties directly from Monte Carlo simulations, we also provide a simple analytic formula that describes their overall behaviour. This facilitates its application to future large datasets and allows for a direct comparison with previous survey parameterisations. For ease of physical interpretation and comparison with other studies, we first convert the linewidths and the corresponding uncertainties from frequency to rest-frame velocity units as follows:

\begin{equation}
\label{eq3}
W_k =  \mathrm{c} (1+z) \frac{W_{\rm f}}{\nu_{\rm 0}},
\end{equation}
where $\mathrm{c}$ is the speed of light, $z$ is the redshift, $W_{\rm f}$ is the barycentric frequency width, and $\nu_{\rm 0}$ is the \HI rest frequency \citep{2017PASA...34...52M}. Using these measurements in velocity units, we express the linewidth uncertainties in fractional form. This formulation is directly applicable to baryonic Tully-Fisher relation analyses, which are typically performed using logarithmic linewidth or rotational velocity. We parameterise the relative linewidth uncertainties separately for the SoFiA and BusyFit measurements by fitting the following power-law relation in logarithmic space,
\begin{equation}
\label{eq4}
\Delta W_{k} / W_{k} = A \, (\mathrm{SNR})^{\alpha} \, W_{k}^{\beta}\, (W_{k,20}-W_{k,50})^{\gamma}.
\end{equation}
Here, $\Delta W_{k}/W_{k}$ is the relative linewidth uncertainty, $W_{k}$ is the linewidth in units of $\mathrm{km/s}$, and $W_{k,20} - W_{k,50}$ serves as a proxy for the profile-edge steepness, also in units of $\mathrm{km\,s^{-1}}$. For simplicity, we omit the subscript $k$ in the remainder of the text, and all linewidth quantities are given in velocity units ($\mathrm{km\,s^{-1}}$). The fitting is performed using ordinary least-squares (OLS) regression implemented in the Python \texttt{statsmodels} package \citep{Seabold2010}, with an iterative $3\sigma$ clipping procedure applied to reduce the influence of outliers. The best-fit parameters for each method and linewidth definition ($\ensuremath{W_{20}}$, $\ensuremath{W_{50}}$, or $\ensuremath{W_{m50}}$) are summarised in Table~\ref{tab:3}. The quoted parameter uncertainties correspond to the $1\sigma$ standard errors. The final column reports the residual scatter of the adopted parameterisation, calculated from the differences between the observed and model-predicted log $\Delta W/W$ values.

\begin{figure*}
\centering
\includegraphics[width=1\linewidth]{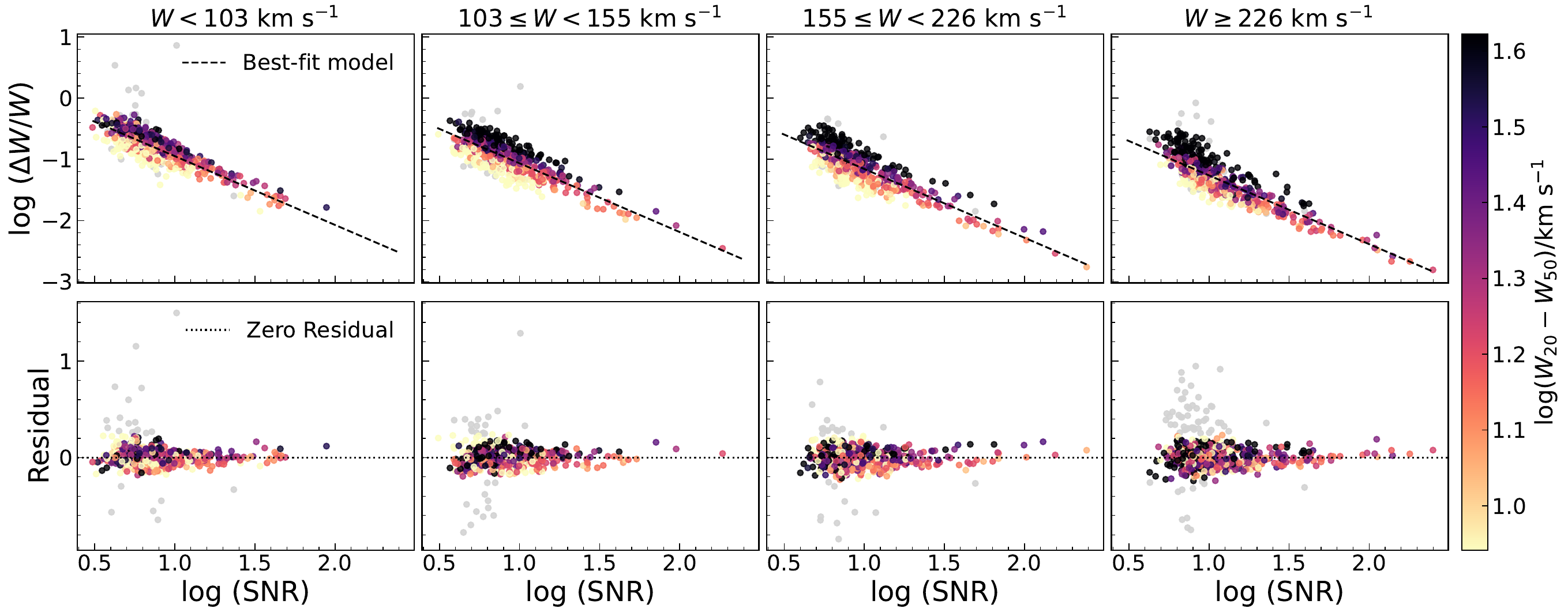}
\caption{Relative linewidth uncertainty scaling relation for the BusyFit $\ensuremath{W_{m50}}$ measurement, shown as a representative example. \textbf{Upper panels:} Logarithmic relative linewidth uncertainty, $\log(\Delta W/W)$, as a function of $\log(\mathrm{SNR})$ for four linewidth bins containing equal numbers of sources in logarithmic linewidth. The dashed lines show the best-fitting model evaluated at the median linewidth and median profile steepness, $W_{20}-W_{50}$, within each linewidth bin, with only SNR allowed to vary. \textbf{Lower panels:} Logarithmic residuals, defined as $\log(\Delta W/W)-\log[(\Delta W/W)_{\rm model}]$, as a function of $\log(\mathrm{SNR})$. In both rows, data points are colour-coded by the logarithmic profile steepness, $\log[(W_{20}- W_{50})/(\mathrm{km\,s^{-1}})]$, while sources excluded by the iterative $3\sigma$-clipping procedure are shown in light grey. The dotted horizontal lines in the lower panels indicate zero residual
($y=0$). The best-fitting parameters for all linewidth definitions are summarised in Table~\ref{tab:3}.}
\label{fig:error_scaling_relation}
\end{figure*}

As a representative example, Figure~\ref{fig:error_scaling_relation} shows the relative uncertainty parameterisation for the BusyFit $\Delta W_{m50}/W_{m50}$. The scaling behaviour and corresponding best-fit parameters are similar to those obtained for the other BusyFit linewidth definitions and for the SoFiA $\Delta W_{50}/ W_{50}$. The lower panels show the logarithmic residuals between the model and the observed uncertainties. Data points are colour-coded by a proxy for profile steepness, defined as log ($W_{20}-W_{50}$), and each column corresponds to a different linewidth bin. The bins are defined in logarithmic space such that each contains an equal number of sources (on average $\sim$520 galaxies per bin across all cases). The dashed lines in the upper panel represent the best-fit model evaluated at the median linewidth and median ($W_{20}-W_{50}$) of each bin, with only SNR allowed to vary. The grey points are data rejected by the $3\sigma$ clipping procedure. The linewidth uncertainties show a strong dependence on all parameters: SNR, profile steepness and linewidth. The SNR power-law indices are close to $-1$ for all BusyFit linewidth definitions and for SoFiA $\ensuremath{W_{50}}$, indicating an approximately inverse scaling. Within each linewidth bin, larger uncertainties are also associated with higher values of ($W_{20}-W_{50}$), corresponding to shallower profile edges. This indicates that the steepness of the spectral profile plays a measurable role in determining the precision of linewidth measurements, because sharper edges allow for a more robust definition of profile boundaries. The above behaviour is consistent with the previous analytical prescriptions \citep{1986AJ.....92..742S,1990A&AS...86..473F} which predict $\Delta W \propto \mathrm{SNR}^{-1} (W_{20}-W_{50})^{1/2}$.

Expressed in fractional form, we find that the relative linewidth uncertainty decreases systematically with increasing linewidth. The fitted linewidth exponent is consistently $\beta$ $\sim$ $-0.6$ to $-0.5$ for all BusyFit linewidth definitions and for SoFiA $\ensuremath{W_{50}}$ measurement. Because \(\Delta W/W\propto W^\beta\), the corresponding absolute uncertainty scales as \(\Delta W\propto W^{\beta+1}\), or approximately \(W^{0.4-0.5}\). Thus, broader profiles have larger absolute uncertainties but smaller fractional uncertainties. The absence of an explicit linewidth dependence in previous studies appears to be due to differing definitions of SNR. Whereas single-dish studies tend to use the channel-to-channel rms as a measure of noise, this measure varies with channel number in interferometric data due to the 3D nature of the mask. We instead use the integrated SNR defined in Equation~\ref{eg:SNR}. The conversion between these SNR definitions is expected to scale approximately as \(W^{0.5}\), which can absorb or expose part of the linewidth dependence in the uncertainty relation.

The SoFiA $\ensuremath{W_{20}}$ and $\ensuremath{W_{m50}}$ show different scaling behaviour from the other linewidth definitions. Their dependence on SNR is generally weaker, with $\alpha$ $\sim$ $-0.6$, while the dependences on linewidth and profile steepness are also reduced. In addition, these relations exhibit larger residual scatter than the other cases, indicating that the adopted parameterisation does not fully capture the behaviour of the SoFiA $\ensuremath{W_{20}}$ and $\ensuremath{W_{m50}}$ uncertainties. Overall, our parameterisation confirms the primary dependence on SNR and profile steepness, identified in previous analytical studies, while demonstrating that an explicit linewidth term is also required to accurately describe the linewidth uncertainties in WALLABY data.

\section{Discussion}
\label{sec:discussion}

\subsection{Measurement comparisons with other surveys}
Whilst our new BusyFit pipeline parameter and uncertainty measurements are a step forward for robust parameterisation and realistic error estimation, there are other sources of non-random errors not considered here. The main one is the definition of the SoFiA mask. If a source is spatially and spectrally resolved, which is almost always the case for WALLABY data (or interferometric data in general), a three-dimensional mask needs to be applied to the data in order to suppress noise and generate a global spectral profile. If some of the \HI emission lies outside the mask, for example at low SNR, then the BusyFit measurements and errors will naturally be affected. Whilst interferometers generally have the advantage of flatter spectral baselines than single dishes, they have the disadvantage that emission may be distributed over many synthesised beams at low SNR, so both spatial and spectral masking are required. 

It is therefore useful to explore the full external error using comparison with independent measurements, preferably with those of higher SNR. Unfortunately, the overlap of the WALLABY pilot survey with large homogeneous surveys is not large at the moment. This will grow as WALLABY coverage increases in equatorial regions. As a step forward, we compare our linewidth measurements with matched measurements from the 2MTF survey (2MASS Tully-Fisher; \citealt{2013MNRAS.432.1178H, 2014MNRAS.443.1044M}) and ALFALFA \citep{2018ApJ...861...49H}. After applying an SNR threshold of 9 to ensure adequate data quality, the resulting cross-matched sample consists of 103 galaxies from ALFALFA and 12 from 2MTF (including 5 observed with the Parkes telescope \citep{2013MNRAS.432.1178H} and 7 with the GBT \citep{2014MNRAS.443.1044M}.

While the WALLABY linewidths are originally measured in the barycentric frequency domain, a direct comparison with ALFALFA and 2MTF requires a consistent velocity scale. We therefore convert the WALLABY linewidths to the rest-frame velocity domain using Equation~\ref{eq3}. Further applications, for example for the Tully-Fisher relation discussed in Section~\ref{sec:BTFR}, will generally involve resolution and dispersion corrections, conversion from linewidth to rotation velocity, and a disk inclination correction. In addition, since the velocity widths reported in the ALFALFA and 2MTF catalogues are given in the optical convention, we convert them to the rest frame by dividing the reported values by (1+$z$). 

\begin{figure*}
\centering
\includegraphics[width=1\linewidth]{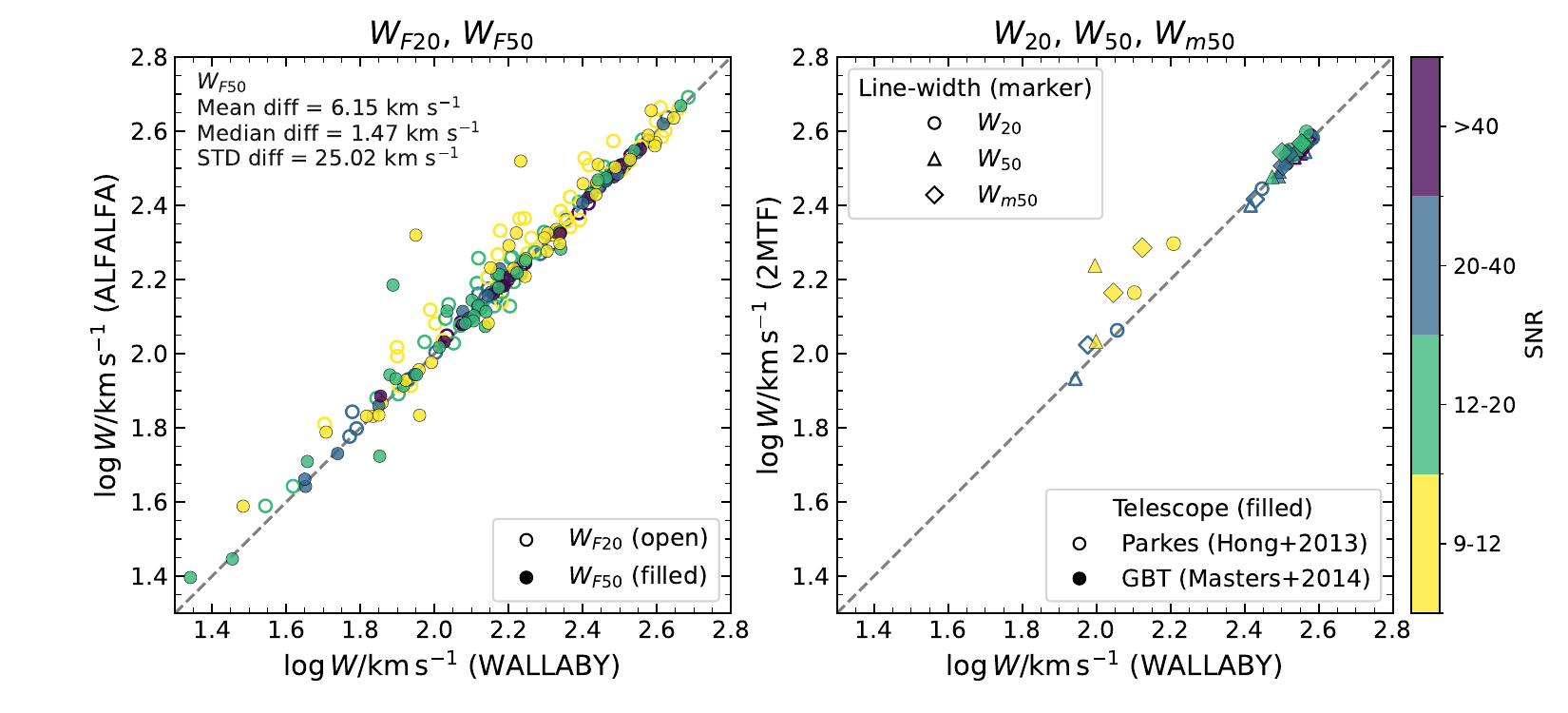}
\caption{\label{fig:ALFALFA_2MTF} Comparison of different definitions of linewidth measurements between WALLABY based on BusyFit model and ALFALFA (left) and 2MTF (right). All ALFALFA and 2MTF linewidths are corrected to the rest-frame. In the left panel, $W_{F20}$ (open circles) and $W_{F50}$ (filled circles) are shown, with the statistical differences in $W_{F50}$ listed in the top-left corner. In the right panel, $\ensuremath{W_{20}}$ (circles), $\ensuremath{W_{50}}$ (triangles), and $\ensuremath{W_{m50}}$ (diamonds) are shown. Measurements from Parkes are indicated by hollow markers, while those from the GBT are shown with filled markers. The points are colour-coded by SNR, and the grey dashed line represents the one-to-one relation. For the matched sample, the mean reported linewidth uncertainties (where available) are 9 $\mathrm{km/s}$ for ALFALFA $W_{F50}$, and 4.4, 3.6, and 4.0 $\mathrm{km/s}$ for the 2MTF Parkes $\ensuremath{W_{20}}$, $\ensuremath{W_{50}}$, and $\ensuremath{W_{m50}}$ measurements, respectively.}
\end{figure*}

Figure~\ref{fig:ALFALFA_2MTF} shows the comparison between our BusyFit measurements and those from ALFALFA (left panel) and 2MTF (right panel). The ALFALFA catalogue quotes $W_{F20}$ and $W_{F50}$, definitions based on both peaks of the double-horn profile, if present (see Section~\ref{sec:3.1}), so we measure the same quantity from the BusyFit models to the WALLABY data. We observe that WALLABY galaxies with lower SNR tend to yield smaller linewidth values compared to their ALFALFA counterparts. As suggested already, this discrepancy may stem from masked flux in the WALLABY spectra and the exclusion of faint high-velocity emission. 
In the 2MTF comparison, we also find that some outliers tend to have a lower SNR. Aside from these outliers, our measurements are generally consistent with the 2MTF results (comparisons with $\ensuremath{W_{20}}$, $\ensuremath{W_{50}}$, and $\ensuremath{W_{m50}}$ are indicated by different markers), with most galaxies following the identity line, save for a few outliers. However, because the overlap between the WALLABY pilot fields and the 2MTF survey are also limited to only a few galaxies, these results—while promising—are not yet statistically conclusive.

\subsection{Application to the baryonic Tully-Fisher relation}
\label{sec:BTFR}

A future application of the techniques presented here is testing the nature of the baryonic Tully-Fisher relation (BTFR) -- namely how fundamental is the BTFR? Does it have intrinsic scatter? How does the intrinsic scatter vary with other parameters such as environment and redshift? Also, is there a linewidth measurement or rotational velocity estimator which minimises that scatter? Finally how do the parameters of the BTFR themselves vary (e.g. slope, zero-point, curvature)?  These have been long-standing questions in BTFR studies and have been investigated in previous works. However, addressing these questions with large \HI surveys requires accurate and homogeneous linewidth measurements, together with realistic estimates of their uncertainties. The linewidth measurement and uncertainty estimation techniques presented here provide a foundation for future BTFR studies with upcoming large \HI surveys. A comprehensive investigation of the BTFR using the full WALLABY sample and the uncertainty framework developed here will be presented in future work.

As a demonstration of how our linewidth measurement and uncertainty framework can be applied to BTFR studies, we use the Hydra DR2 sample from WALLABY PDR1. We further restrict the sample based on visual inspection of the \HI\ moment maps and optical Digitized Sky Survey (DSS) images, excluding galaxies showing signs of interaction or merging, paired galaxies, those only partially detected in \HI by SoFiA, or those with strong continuum artefacts. We then apply the following selection criteria: SNR $>$ 8; inclination $>$ 45 deg to minimise the impact of inclination uncertainties on deprojected rotational velocities; and linewidth measurement uncertainties $< 30$ km/s. Profiles for which BusyFit failed to converge are also excluded, resulting in a final sample size of 84 galaxies.

The baryonic mass is defined as
\begin{equation}
M_{\rm bar} = M_\star + M_{\rm gas},
\end{equation}
where $M_\star$ and $M_{\rm gas}$ are the stellar and gas masses, respectively. The stellar mass is calculated using the WISE band 1 (W1) magnitude from \citet{2023MNRAS.519.4589C}, obtained via the photometric pipeline of \citet{2013AJ....145....6J}, following the mass-to-light formulation of \citet{2023ApJ...946...95J}. The uncertainty on the stellar mass is derived by propagating the photometric uncertainties in the WISE W1 magnitude. The gas mass is defined as
\begin{equation}
M_{\rm gas} = 1.33 \, M_{\rm HI},
\end{equation}
where the factor of 1.33 accounts for the contribution of helium and metals. We neglect the molecular gas, as it typically contributes only a small fraction of the total baryonic mass in nearby galaxies (e.g. \citealt{2015ApJ...802...18M, 2018MNRAS.474.4366P}). The \HI mass is derived from the integrated \HI flux in the SoFiA catalog. Before conversion to \HI mass, the integrated \HI flux is statistically corrected following \citet{2022PASA...39...58W} to account for the flux deficit relative to ALFALFA. The uncertainty in the \HI mass is propagated from the integrated \HI flux uncertainties reported in the SoFiA catalogue, including the additional systematic correction factor. To approximate the maximum rotational velocity, the velocity width is divided by 2 and corrected for inclination following the equation:

\begin{equation}
    V_{\textrm{max}} = \frac{W_{\textrm{k}}}{2 \ \textrm{sin}(i)},
\end{equation}
where $W_{\textrm{k}}$ is linewidth translated into the rest-frame velocity domain (see Eq.~\ref{eq3}), and $i$ is inclination measured from WISE W1 band (\citealt{2023MNRAS.519.4589C}). As noted by \citet{2023MNRAS.519.4589C}, these inclinations may be underestimated because the limited spatial resolution of W1 imaging and its sensitivity to older stellar populations, particularly the bulge, can result in rounder axis ratios. The uncertainty in the rotational velocity is obtained via standard error propagation, including contributions from both linewidth and inclination uncertainties.

Since not all galaxies in the Hydra DR2 are members of the Hydra cluster, we classify cluster membership based on the projected phase-space diagram following \citet{2017ApJ...843..128R}. Galaxies are classified as cluster members if their projected distance from the cluster centre is within $3 R_{200}$ and their velocity relative to the cluster systematic velocity is less than the escape velocity. These galaxies are assumed to be at the cluster distance of 47.5 $\pm$ 3~Mpc \citep{2017ApJ...843...16K}. Otherwise, they are classified as field galaxies, with distances derived from Cosmicflows-3 \citep{2020AJ....159...67K} via a velocity field reconstruction based on a Wiener filter model \citep{2019MNRAS.488.5438G}. We adopt a conservative distance uncertainty of 15\% for field galaxies, following the distance-accuracy threshold used in the construction of the Cosmicflows-3 model. For both samples, these distance uncertainties are propagated into the stellar and gas mass estimates, contributing to the total baryonic mass uncertainty budget.

\begin{figure*}
\centering
\includegraphics[width=1\linewidth]{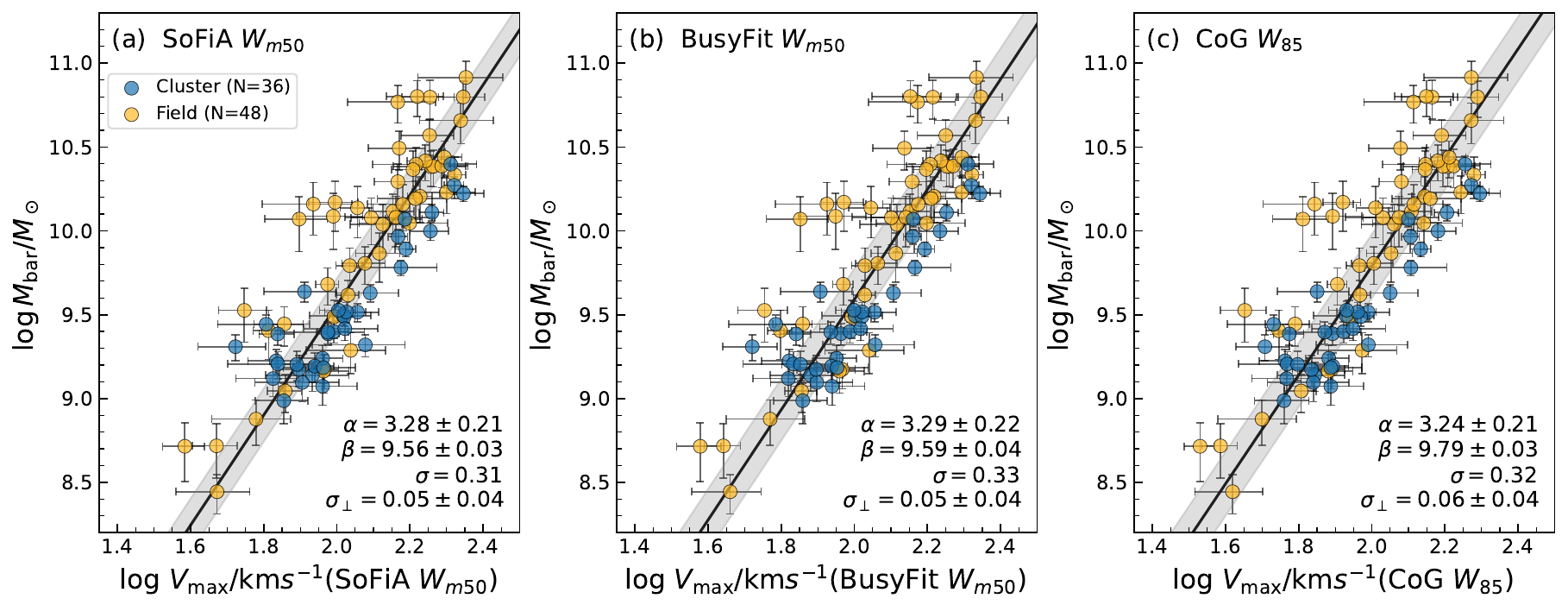}
\caption{The baryonic Tully-Fisher relations constructed using three different width measurements: SoFiA $\ensuremath{W_{m50}}$ (left), BusyFit $\ensuremath{W_{m50}}$ (middle), and CoG $\ensuremath{W_{85}}$ (right). The black solid line shows the best-fitting relations, and the 1$\sigma$ uncertainty of the fit, derived from MCMC posterior samples (see Figure~\ref{fig:TFR_MCMC}), is shown with a grey shaded region. The corresponding best-fit parameters are listed in the bottom right corner of each plot. Cluster galaxies are shown in blue and field galaxies in orange; the number of galaxies in each subsample is given in parentheses. For cluster galaxies, a fixed distance of 47.5 Mpc is adopted, while for field galaxies distances are obtained from the Cosmicflows-3 model.}
\label{fig:TFR}
\end{figure*}

\begin{figure}
\centering  
\includegraphics[width=\linewidth]{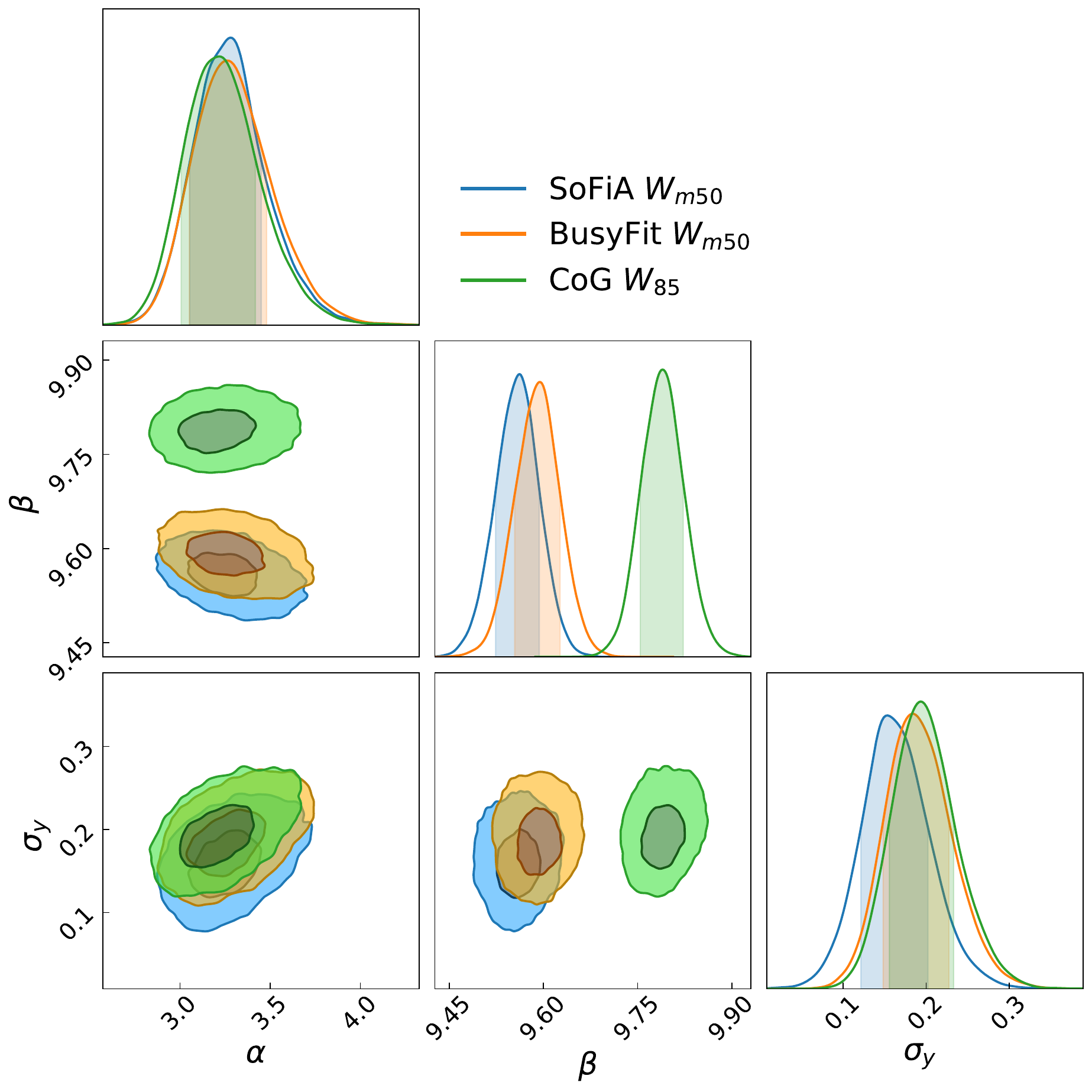}
\captionsetup{justification=raggedright, singlelinecheck=false} 
\caption{\label{fig:TFR_MCMC} MCMC posterior distributions of the BTFR fit parameters obtained using different linewidth measurements. Here, $\alpha$ denotes the slope, $\beta$ the intercept at $x=2.0$, and $\sigma_y$ the intrinsic vertical scatter. Contours show the 1$\sigma$ and 2$\sigma$ credible regions.}
\end{figure}

The constructed BTFR is presented in Figure~\ref{fig:TFR} using three different linewidth definitions: panel (a) SoFiA $\ensuremath{W_{m50}}$, panel (b) BusyFit $\ensuremath{W_{m50}}$, and panel (c) CoG $\ensuremath{W_{85}}$. To quantify this relation, we fit the data using HyperFit (\citealt{2015PASA...32...33R}), which minimises orthogonal residuals and derives statistical parameters and their uncertainties via MCMC sampling. The best-fit line is shown as a solid black line, with the grey shaded region representing the 1 $\sigma$ intrinsic vertical scatter inferred from the MCMC posterior samples (Figure~\ref{fig:TFR_MCMC}). The best-fit parameters are provided in each panel and tabulated in Table~\ref{tab:tfr_results}. Here, $\alpha$ represents the slope of the best-fit line, $\beta$ is the value of $y$ at $x = 2.0$, $\sigma$ is the observed vertical scatter, $\sigma_y$ is the intrinsic vertical scatter, and $\sigma_\perp$ is the orthogonal intrinsic scatter. 

Considering the statistical uncertainties, the BTFR parameters do not differ significantly between the different linewidth measurements, with the exception of $\beta$. Cluster members in the high-mass regime tend to lie below the best-fit line, which may be a result of environmental effects, selection effects, or distance errors. However, robust conclusions are difficult to draw due to the limited sample size. We will investigate these effects in more detail in a subsequent paper using the full WALLABY survey sample.

\begin{table}
\centering
\large
\renewcommand{\arraystretch}{1.8}
\setlength{\tabcolsep}{4pt}

\resizebox{\columnwidth}{!}{
\begin{tabular}{lcccc}
\hline
\hline
Method & $\alpha$ (slope) & $\beta$ (intercept at x$=$2) & $\sigma$ & $\sigma_\perp$ \\
\hline
SoFiA $\ensuremath{W_{m50}}$ & 3.28 $\pm$ 0.21 & 9.56 $\pm$ 0.03 & 0.31 & 0.05 $\pm$ 0.04 \\
BusyFit $\ensuremath{W_{m50}}$ & 3.29 $\pm$ 0.22 & 9.59 $\pm$ 0.04 & 0.33 & 0.05 $\pm$ 0.04 \\
CoG $\ensuremath{W_{85}}$ & 3.24 $\pm$ 0.21 & 9.79 $\pm$ 0.03 & 0.32 & 0.06 $\pm$ 0.04 \\
\hline
\end{tabular}
}
\caption{Best-fitting parameters of the BTFR for different linewidth measurement methods.}
\label{tab:tfr_results}
\end{table}

\section{Summary and conclusion}
\label{sec:summary}


We have analysed the \HI global profiles from the WALLABY pilot data using the spectral profile fitting tool, BusyFit, and developed a robust and automated pipeline for linewidth and uncertainty measurements. By comparing several fitting approaches with BusyFit, we identified a method that reliably recovers linewidths across a wide range of SNR ranges, particularly in the low-SNR regime typical of WALLABY data. 

Our findings demonstrate that parameterising spectral profiles using BusyFit provides more robust and accurate linewidths compared to direct measurement methods such as SoFiA. Based on the comparison between SoFiA and BusyFit linewidth measurements results (Figure~\ref{fig:Compare_SoFiA} and Table~\ref{tab:line_width_statistics}), we demonstrate that $\ensuremath{W_{20}}$ and $\ensuremath{W_{m50}}$ are inherently robust measurements. Specifically, our Monte Carlo simulations (Figure~\ref{fig:6}) identify $\ensuremath{W_{m50}}$ as the most stable metric, exhibiting the smallest deviations between the model linewidths and median values derived from mock spectra.

We further establish an empirical parameterisation of linewidth uncertainties, confirming that they scale strongly with the integrated SNR (approximately inversely), profile steepness, and the linewidth. This highlights that measurement uncertainties are governed not only by data quality but also by the intrinsic properties of the spectral profile, including its width and morphology.

Although $\ensuremath{W_{m50}}$ is widely used as a proxy for the maximum rotational velocity of galaxies in Tully-Fisher surveys (e.g. \citealt{2009AJ....138.1938C}), it remains challenging to determine whether it provides the best representation of the underlying rotational velocity. As shown in Figure~\ref{fig:TFR}, the current sample size is insufficient to statistically distinguish between different linewidth definitions. Moreover, uncertainties in inclination corrections introduce significant scatter, making it difficult to isolate the intrinsic scatter of the BTFR and to definitively rank the performance of different linewidth metrics in a cosmological context. These questions will be revisited in future work using the substantially larger sample provided by the full WALLABY survey. 

\section*{Acknowledgements}
We are grateful to other WALLABY team members for their helpful comments on an earlier version of the manuscript. Parts of this research were supported by the Australian Research Council Centre of Excellence for All Sky Astrophysics in 3 Dimensions (ASTRO 3D), through project number CE170100013.
DO is a recipient of an Australian Research Council Future Fellowship (FT190100083) funded by the Australian Government. Matthew Colless was funded by the Australian Research Council (https://ror.org/05mmh0f86) under Discovery Project grant DP260101015. HMC acknowledges support from the Institut Universitaire de France. B. L. was supported by the National Research Foundation of Korea (NRF) grant funded by the Korea goverment (MSIT) (RS-2026-25498235). This scientific work uses data obtained from Inyarrimanha Ilgari Bundara, the CSIRO Murchison Radio-astronomy Observatory. We acknowledge the Wajarri Yamaji People as the Traditional Owners and native title holders of the Observatory site. CSIRO’s ASKAP radio telescope is part of the Australia Telescope National Facility (\url{https://ror.org/05qajvd42}). Operation of ASKAP is funded by the Australian Government with support from the National Collaborative Research Infrastructure Strategy. ASKAP uses the resources of the Pawsey Supercomputing Research Centre. Establishment of ASKAP, Inyarrimanha Ilgari Bundara, the CSIRO Murchison Radio-astronomy Observatory and the Pawsey Supercomputing Research Centre are initiatives of the Australian Government, with support from the Government of Western Australia and the Science and Industry Endowment Fund. This work was supported by the Australian SKA Regional Centre (AusSRC), Australia’s portion of the international SKA Regional Centre Network (SRCNet), funded by the Australian Government through the Department of Industry, Science, and Resources (DISR; grant SKARC000001). AusSRC is an equal collaboration between CSIRO – Australia’s national science agency, Curtin University, the Pawsey Supercomputing Research Centre, and the University of Western Australia.

\section*{Data availability}
The WALLABY source catalogue and associated data products (e.g. integrated spectra) are publicly available through the CSIRO ASKAP Science Data Archive (CASDA) and the Canadian Astronomy Data Centre (CADC). Information on data access, software tools used to generate the data products, and documentation including example scripts for accessing the data are available via the WALLABY Data Portal \url{(https://wallaby-survey.org/data/)}.



\bibliographystyle{mnras}
\bibliography{example} 





\bsp	
\label{lastpage}
\end{document}